\documentclass[twocolumn,twocolappendix]{aastex631}

\usepackage{amsmath}
\usepackage{soul}
\definecolor{dgreen}{RGB}{26,148,49}
\definecolor{forest}{RGB}{3, 148, 49}
\newcommand{\YT}[1]{\textcolor{red}{[{\bf YT}: #1]}}

\shorttitle{Bayesian Multi-line Intensity Mapping}
\shortauthors{Cheng et al.}
\graphicspath{{./}{\YT{figure}s/}}

\begin{document}

\title{Bayesian Multi-line Intensity Mapping}

\author[0000-0002-5437-0504]{Yun-Ting Cheng}
\address{California Institute of Technology, 1200 E. California Boulevard, Pasadena, CA 91125, USA}
\address{Jet Propulsion Laboratory, California Institute of Technology, 4800 Oak Grove Drive, Pasadena, CA 91109, USA}\email{ycheng3@caltech.edu}

\author[0009-0000-4756-4223]{Kailai Wang}
\address{Department of Physics, Cornell University, 616 Thurston Avenue, Ithaca, NY 14853, USA}

\author[0000-0002-5854-8269]{Benjamin D. Wandelt}
\address{Sorbonne Universit{\'e}, CNRS, UMR 7095, Institut d’Astrophysique de Paris, 98 bis bd Arago, 75014 Paris, France}
\address{Center for Computational Astrophysics, Flatiron Institute, 162 5th Avenue, New York, NY 10010, USA}

\author[0000-0001-5929-4187]{Tzu-Ching Chang}
\address{Jet Propulsion Laboratory, California Institute of Technology, 4800 Oak Grove Drive, Pasadena, CA 91109, USA}
\address{California Institute of Technology, 1200 E. California Boulevard, Pasadena, CA 91125, USA}

\author[0000-0001-7432-2932]{Olivier Dor{\'e}}
\address{Jet Propulsion Laboratory, California Institute of Technology, 4800 Oak Grove Drive, Pasadena, CA 91109, USA}
\address{California Institute of Technology, 1200 E. California Boulevard, Pasadena, CA 91125, USA}

\begin{abstract}
Line intensity mapping (LIM) has emerged as a promising tool for probing the 3D large-scale structure through the aggregate emission of spectral lines. The presence of interloper lines poses a crucial challenge in extracting the signal from the target line in LIM. In this work, we introduce a novel method for LIM analysis that simultaneously extracts line signals from multiple spectral lines, utilizing the covariance of native LIM data elements defined in the spectral--angular space. We leverage correlated information from different lines to perform joint inference on all lines simultaneously, employing a Bayesian analysis framework. We present the formalism, demonstrate our technique with a mock survey setup resembling the SPHEREx deep field observation, and consider four spectral lines within the SPHEREx spectral coverage in the near infrared: H$\alpha$, $[$\ion{O}{3}$]$, H$\beta$, and $[$\ion{O}{2}$]$. We demonstrate that our method can extract the power spectrum of all four lines at the $\gtrsim 10\sigma$ level at $z<2$. For the brightest line, H$\alpha$, the $10\sigma$ sensitivity can be achieved out to $z\sim3$. Our technique offers a flexible framework for LIM analysis, enabling simultaneous inference of signals from multiple line emissions while accommodating diverse modeling constraints and parameterizations.
\end{abstract}


\section{Introduction} \label{S:intro}
Line intensity mapping \citep[LIM; for reviews, see][]{2017arXiv170909066K, 2022A&ARv..30....5B} is an emerging technique for studying the large-scale structure (LSS) of the Universe. By mapping the emission from specific spectral lines and determining their redshifts from observed frequencies, LIM traces the three-dimensional (3D) LSS using cumulative emissions from all sources. It serves as a promising method for bridging the gap in LSS probes between the recombination era explored by the cosmic microwave background and the lower-redshift Universe ($z\lesssim3$) accessible with current and upcoming galaxy surveys---e.g.,  Sloan Digital Sky Survey \citep{2006PhRvD..74l3507T}, Dark Energy Survey \citep{2018PhRvD..98d2006E, 2022PhRvD.105b3520A}, Dark Energy Spectroscopic Instrument \citep{2016arXiv161100036D}, Euclid \citep{2011arXiv1110.3193L}, the Rubin Observatory \citep{2009arXiv0912.0201L}, SPHEREx \citep{2014arXiv1412.4872D, 2016arXiv160607039D, 2018arXiv180505489D}, and the Nancy Grace Roman Space Telescope \citep{2015arXiv150303757S}. Additionally, LIM provides crucial constraints on the collective properties of the interstellar medium (ISM) of galaxies across cosmic time through the measurement of aggregate line emission.

The field of LIM, initially pioneered by 21 cm cosmology with a primary focus on probing the Dark Ages and the Epoch of Reionization \citep[EoR;][]{2006PhR...433..181F,2010ARA&A..48..127M,2012RPPh...75h6901P}, has since expanded to include atomic or molecular lines across a broad electromagnetic spectrum. In the submillimeter wavelengths, several LIM experiments targeting $[$\ion{C}{2}$]$ and/or CO rotational ladders have reported preliminary detections or provided upper-limit constraints, including COPSS \citep{2015ApJ...814..140K,2016ApJ...830...34K}, mmIME \citep{2020ApJ...901..141K}, and COMAP \citep{2022ApJ...933..182C,2022ApJ...933..188B}, with more measurements anticipated from ongoing and upcoming experiments like FYST \citep{2023ApJS..264....7C}, CONCERTO \citep{2020A&A...642A..60C}, TIME \citep{2014SPIE.9153E..1WC,2021ApJ...915...33S}, SPT-SLIM \citep{2022JLTP..209..758K}, EXCLAIM \citep{2020JLTP..199.1027A}, and TIM \citep{2020arXiv200914340V}. In the optical and near-infrared regimes, the ongoing HETDEX experiment \citep{2008ASPC..399..115H,2021ApJ...923..217G} is conducting Ly$\alpha$ LIM at $z\sim2-4$. The upcoming near-infrared all-sky spectral survey SPHEREx will explore emissions from multiple lines, including H$\alpha$, $[$\ion{O}{3}$]$, H$\beta$, and $[$\ion{O}{2}$]$, among others \citep{2023arXiv231204636F}. Last, the proposed next-generation far-infrared observatories, such as PRIMA \citep{2023arXiv231020572M}, are poised to conduct LIM across various far-infrared lines ($[$\ion{Ne}{2}$]$, $H_2$, $[$\ion{S}{3}$]$, $[$\ion{Si}{2}$]$, etc.).

The spectral coverage of most of the LIM experiments, except for 21 cm LIM, are accessible to multiple spectral lines from different redshift ranges. While analyzing signals from multiple lines has the potential to unveil more ISM physics, as different lines often trace different ISM environments and/or dust properties, it also poses a significant data analysis challenge, since the emissions from different lines are mixed in the LIM dataset.

Extensive studies have explored different strategies for tackling the challenge of line confusion in LIM. One approach involves masking voxels---fundamental 3D LIM data elements defined by pixels and spectral channels---that contain bright interlopers, using external galaxy survey catalogs to mitigate the interloper line signal \citep{2015MNRAS.450.3829Y,2015ApJ...806..209S,2018ApJ...856..107S,2022A&A...667A.156B,2023A&A...676A..62V}. External catalogs can also be used in cross-correlation with LIM data, enabling the extraction of signals from individual lines \citep{2013ApJ...763..132S,2015ApJ...806..209S,2016MNRAS.457.3541C,2019ApJ...872..186C,2013ApJ...768...15P,2019MNRAS.489L..53Y,2023ApJ...956...84V}. Furthermore, with the LIM dataset, different line signals can also be distinguished by the anisotropy of the interloper power spectrum arising from projection to the comoving frame of the target line \citep{2016ApJ...825..143L,2016ApJ...832..165C,2020ApJ...894..152G}.

Another approach for mitigating line confusion in LIM involves leveraging the correlation between pairs of observed frequencies containing different lines from the same redshift. This correlation arises because these frequencies trace the same underlying LSS, while the interlopers in each channel of the pair are uncorrelated, originating from distinct line-of-sight (LOS) distances. \citet{2020ApJ...901..142C} utilize this information to extract an intensity map from individual lines in LIM using a spectral-template-fitting technique. In addition, previous studies have explored the possibility of probing cross-spectra between two different lines within the same or different LIM datasets \citep{2010JCAP...11..016V,2011JCAP...08..010V,2012ApJ...745...49G,2016ApJ...833..153S,2023arXiv231208471R}. Furthermore, some proposed methods use combinations of multiple cross-line power spectra to estimate the autospectrum of individual lines \citep{2019ApJ...874..133B,2021JCAP...05..068S,2023arXiv230800749M}. The cross-bispectrum of two lines can also achieve similar results in the same spirit \citep{2018ApJ...867...26B}.

In this study, we present a novel method for simultaneously extracting the line signal from multiple spectral lines in a LIM dataset, building upon the technique introduced in \citet[][hereafter, C23]{2023ApJ...944..151C}. In C23, we developed an inference framework to reconstruct the LSS clustering, the spectral energy distribution (SED), and the LOS distribution of emitting sources from the cross-frequency angular power spectra, $C_{\ell,\nu\nu'}$'s. C23 demonstrated the effectiveness of this technique by applying it to multiple broadband photometric maps. C23 found that sharp features in the SED significantly enhance parameter constraints, since they break the degeneracy between spectral and redshift information in 2D frequency maps. Given this finding, LIM emerges as a promising application of the technique, as line emissions are inherently sharp features in the SED. Therefore, this study aims to extend the method developed in C23 to LIM to reconstruct the redshift evolution of line emissions. In LIM, the 3D spatial distribution of line emission, which traces the underlying LSS, is encoded in 3D spectral--angular space. On large scales, all two-point-level information in a LIM dataset is contained in the data covariance, or equivalently, in harmonic space, the cross-frequency angular power spectra $C_{\ell,\nu\nu'}$'s with all combinations of observed frequency channels $\nu$ and $\nu'$. By assuming only the homogeneity and isotropy of cosmological density fluctuations, along with knowledge of the rest-frame frequencies of spectral lines, we employ a Bayesian approach to extract the bias-weighted line intensity as a function of redshift for all spectral lines in LIM from the data covariance, $C_{\ell,\nu\nu'}$.

In contrast to many aforementioned methods for separating spectral lines in LIM, our approach avoids the common step of projecting the LIM data into 3D comoving space at an assumed central redshift. Instead, we directly model the covariance in the native data space, i.e., in the spectral-angular space, where the data covariance ($C_{\ell,\nu\nu'}$) naturally incorporates anisotropic projection, line-line correlations, and the redshift evolution of line intensity and cosmological fluctuations within our formalism.

To demonstrate our method, we implement it in a simulated survey resembling the SPHEREx deep-field configuration with its expected sensitivity, and consider four spectral lines--- H$\alpha$, $[$\ion{O}{3}$]$, H$\beta$, and $[$\ion{O}{2}$]$---within the SPHEREx spectral coverage. We apply our algorithm to a simulated observed data covariance ($C_{\ell,\nu\nu'}$) and assess the uncertainties with our inference of the input line signals.

This paper is organized as follows. Section~\ref{S:modeling} provides a detailed description of the LIM signal and power spectra. Our assumed survey setup and the models for line signals are outlined in Sections~\ref{S:survey_setup} and \ref{S:line_modeling}, respectively. Section~\ref{S:algorithm} introduces our inference algorithm. The results of applying our technique to mock observed data are presented in Section~\ref{S:results}, followed by additional discussions in Section~\ref{S:discussion}. We highlight the unique advantages of our method in Section~\ref{S:unique_advantages} and provide discussions on comparisons with relevant previous works in Section~\ref{S:compare_previous_works}. Future prospects for extending this work are discussed in Section~\ref{S:future_works}, and the conclusion is provided in Section~\ref{S:conclusion}. Throughout this work, we assume a flat $\Lambda$CDM cosmology consistent with the measurements from Planck \citep{2020AA...641A...6P}.

\section{Power Spectrum Modeling}\label{S:modeling}
In this section, we present the formalism for the intensity field from line emission in LIM in Section~\ref{S:intensity} and for the auto- and cross-frequency angular power spectra, $C_{\ell}^{\nu\nu'}$, in Section~\ref{S:Cl_nunu1_formalism}. Only the main expressions are presented here; more detailed derivations are provided in Appendix~\ref{A:intensity_and_PS}.

\subsection{Intensity Field}\label{S:intensity}
The intensity field at an observed frequency $\nu$ and angular position $\hat{n}$ is given by the emission from continuum, spectral lines, foregrounds, and the noise:
\begin{equation}
\begin{split}
\nu I_\nu(\nu, \hat{n}) &= \nu I_\nu^{\rm cont}(\nu, \hat{n}) + \nu I_\nu^{\rm line}(\nu, \hat{n}) \\
& + \nu I_\nu^{\rm FG}(\nu, \hat{n}) + \nu I_\nu^{ n}(\nu, \hat{n}).
\end{split}
\end{equation}
The continuum and the foreground usually have a smooth spectrum, which can be effectively mitigated by methods that filter out the smooth spectral component in the data (see Section~\ref{S:continuum} for discussions)\footnote{We note that some foregrounds, like Galactic and atmospheric foregrounds, may contain spectral line emission or absorption. Thus, in practice, additional data-processing steps may be necessary to remove these foreground line features before implementing our analysis.}. The line signal might be altered during the high-pass-filtering process in foreground cleaning, and we assume this potential bias has already been corrected in this study. However, we emphasize that any line signal transfer function induced by the foreground filtering must be carefully characterized in practice.

Therefore, in this work, we assume that the LIM dataset only contains the line emission from the $N_{\rm line}$ number of spectral lines and the noise fluctuations:
\begin{equation}
\begin{split}
\nu I_\nu(\nu, \hat{n})=&\nu I_\nu^{\rm line}(\nu, \hat{n}) + \nu I_\nu^{n}(\nu, \hat{n})\\
=&\sum_{i=1}^{N_{\rm line}}\nu I_\nu^i(\nu, \hat{n}) + \nu I_\nu^{n}(\nu, \hat{n}).
\end{split}
\end{equation}

In typical LIM experiments, the spectral resolution is not sufficient to resolve the intrinsic line profile from sources, and thus we model the line profile as a Dirac delta function at the rest-frame line frequency $\nu_{\rm rf}^i$, which gives the line intensity in the following form (see Appendix~\ref{A:line_intensity} for detailed derivations):
\begin{equation}\label{E:nuInu_line}
\nu I_\nu^i(\nu, \hat{n}) = \frac{c(1+z_{i\nu})}{H(z_{i\nu})}M_{0,i}(\chi_{i\nu}, \hat{n}) A_0(\chi_{i\nu}),
\end{equation}
where $c$ is the speed of light, $H(z)$ is the Hubble parameter, and $z_{i\nu} = \nu_{\rm rf}^i/\nu - 1$ is the redshift of the $i$th line at the observed frequency\footnote{Throughout this manuscript, $\nu$ is referred to as the observed frequency, and the rest-frame frequency is denoted by $\nu_{\rm rf}$.} $\nu$. We define $\chi_{i\nu}$ as the comoving distance at redshift\footnote{The redshift $z$ and the comoving distance $\chi$ will be used interchangeably to describe the LOS distance.} $z_{i\nu}$, and
$M_{0,i}(\chi, \hat{n})= dL_i(\chi,\hat{n})/dV$ is the comoving line luminosity density.

On large scales, $M_{0,i}(\chi, \hat{n})$ follows the underlying matter density field $\delta_m(\chi, \hat{n})$ with a scale-independent luminosity-weighted bias factor $b_i(\chi)$:
\begin{equation}
M_{0,i}(\chi, \hat{n}) = b_i(\chi)M_{0,i}(\chi)[1+\delta_m(\chi , \hat{n})],
\end{equation}
where $M_{0,i}(\chi)$ is the mean luminosity density averaged over angular positions $\hat{n}$, and
\begin{equation}
A_0(\chi) = D_A^2(\chi)/4\pi D_L^2(\chi),
\end{equation}
where $D_A$ and $D_L$ are the comoving angular diameter distance and luminosity distance, respectively.

We model the noise at the frequency band $\nu$ as a zero-mean Gaussian fluctuation with the variance $\sigma_n^2(\nu)$:
\begin{equation}
\nu I_\nu^{n}(\nu, \hat{n}) \sim \mathcal{N}(0, \sigma^2_n(\nu)).
\end{equation}
We assume the noise in each frequency channel is independent, and thus there is no cross-channel noise covariance.

\subsection{Angular Power Spectrum}\label{S:Cl_nunu1_formalism}
We describe the covariance of the LIM dataset in terms of the auto- and cross-angular power spectra, $C_{\ell,\nu\nu'}$'s, of all combinations of the frequency channels $\nu$ and $\nu'$. On large scales, ignoring the redshift space distortion (RSD) effect, the line emission field is isotropic, and its fluctuations can be fully described by a Gaussian probability distribution. Therefore, the $C_{\ell,\nu\nu'}$'s capture the full two-point information from the LIM dataset on large scales \citep{2013acna.conf...87W}. In this work, we focus only on two-point statistics. However, we note that the LIM field on small scales is highly non-Gaussian, and the power spectrum alone is insufficient to capture the full information. This has motivated previous studies to include one-point statistics to exploit more information from LIM data \citep{2019ApJ...871...75I,2022arXiv220901223B,2023MNRAS.520.5305C}.

The total power spectrum is the sum of the contribution from emission lines and noise:
\begin{equation}\label{E:total_Cl}
\mathbf{C}_{\ell} = \mathbf{C}^{\rm line}_{\ell} + \mathbf{C}^{n}_{\ell} = \sum_{i=1}^{N_{\rm line}} \sum_{i'=1}^{N_{\rm line}} \mathbf{C}_{\ell,ii'} + \mathbf{C}^{n}_{\ell}.
\end{equation}
Here, the boldface $\mathbf{C}_{\ell}$ denotes the angular-power-spectrum matrix of size $N_{\nu}\times N_{\nu}$, where $N_{\nu}$ is the total number of spectral channels, and each element of $\mathbf{C}_{\ell}$ is given by $C_{\ell,\nu\nu'}$.

In this work, we only consider information from large (linear) scales, ignoring fluctuations from nonlinear clustering and Poisson noise. We verify that, for the scales considered in this work, the Poisson noise power is negligible (Appendix~\ref{A:SN}). Therefore, the angular power spectrum from emission lines can be expressed as
\begin{equation}\label{E:Cl_line_bessel}
\begin{split}
C^{\rm line}_{\ell,\nu\nu'} &= \sum_{i=1}^{N_{\rm line}} \sum_{i'=1}^{N_{\rm line}}C_{\ell,\nu\nu'ii'}\\
&= \sum_{i=1}^{N_{\rm line}} \sum_{i'=1}^{N_{\rm line}}\int d\chi W_{i\nu}(\chi) \int d\chi' W_{i'\nu'}(\chi') \\
&\cdot\int \frac{dk}{k}\frac{2}{\pi} k^3 P(k) D(\chi) D(\chi') j_\ell(k\chi) j_\ell(k\chi'),
\end{split}
\end{equation}
where $D(\chi)$ is the linear growth factor, and $P(k)$ is the linear matter power spectrum at the present time. $W_{i\nu}(\chi)$ is the window function of the frequency channel $\nu$ for the emission from line $i$. Here, we assume the observing filter profile is a narrow top-hat function centered at the observed frequency $\nu$ with a width $\Delta \nu$ that spans the frequency range $\nu_{\rm min} < \nu < \nu_{\rm max}$. With this assumption, the window function can be expressed as (see Appendix~\ref{A:window_function} for detailed derivations)
\begin{equation}
\begin{split}
W_{i\nu}(\chi) &=
\begin{cases}
\frac{\nu}{\Delta\nu}b_{i}(\chi) M_{0,i}(\chi) A_{0}(\chi)
& \quad\text{if } \chi^{\rm min}_{i\nu} < \chi < \chi^{\rm max}_{i\nu}\\
0 & \quad\text{otherwise},
\end{cases}
\end{split}
\end{equation}
where $\chi^{\rm min/max}_{i\nu} = \chi(z^{\rm min/max}_{i\nu})$, and $z^{\rm min/max}_{i\nu}=(\nu^i_{\rm rf}/\nu^{\rm max/min})-1$. $z^{\rm min/max}_{i\nu}$ and $\chi^{\rm min/max}_{i\nu}$ denote the corresponding redshift and comoving distance that can be probed by the $i$th line in the frequency channel $\nu$ spanning the frequency range $\nu_{\rm min} < \nu < \nu_{\rm max}$.

As most current and upcoming LIM experiments target a relatively small field size (typically around degree scales), we adopt the Limber approximation \citep{1953ApJ...117..134L}, which is valid for small survey sizes \citep[e.g.,][]{2013MNRAS.432.2945H}. The angular power spectrum can thus be simplified as
\begin{equation}\label{E:Cl_line}
\begin{split}
C_ {\ell,\nu\nu'ii'} =  \int \frac{d\chi}{\chi^2} W_{i\nu}(\chi) W_{i'\nu'}(\chi) D^2(\chi) P(\frac{\ell+\frac{1}{2}}{\chi}).
\end{split}
\end{equation}
We define 
\begin{equation}
M_i(\chi) = b_i(\chi)M_{0,i}(\chi).
\end{equation}
Hereafter, the term ``bias-weighted luminosity'' refers to $M_i(\chi)$ and similarly, ``bias-weighted intensity'' refers to the quantity $b_i(\chi)\nu I_\nu^i(\chi)$. We also define
\begin{equation}
A(\chi) = D(\chi)A_{0,i}(\chi).
\end{equation}
Then, the angular power spectrum from the lines can be expressed as
\begin{equation}\label{E:Cl_nunu1ii1_limber}
\begin{split}
C_ {\ell,\nu\nu'ii'} &= \frac{\nu}{\Delta\nu}\frac{\nu'}{\Delta\nu'}\frac{\Delta\chi_{ii'\nu\nu'}}{\chi^2_{ii'\nu\nu'}}A^2(\chi_{ii'\nu\nu'})\\
&\cdot M_i(\chi_{ii'\nu\nu'})M_{i'}(\chi_{ii'\nu\nu'})P(\frac{\ell+\frac{1}{2}}{\chi_{ii'\nu\nu'}}),
\end{split}
\end{equation}
where $\chi_{ii'\nu\nu'}$ and $\Delta \chi_{ii'\nu\nu'}$ denote the center and the width of the overlapping comoving distance between $\chi^{\rm min}_{i\nu}<\chi<\chi^{\rm max}_{i\nu}$ and $\chi^{\rm min}_{i'\nu'}<\chi<\chi^{\rm max}_{i'\nu'}$, respectively. For the autospectra of a line ($i=i'$), if there is no overlap between the filter profile, which is the case we consider in this work, the $C^{\rm line}_ {\ell,\nu\nu'ii'}$'s are only nonzero for the same spectral channel (i.e., $\nu=\nu'$). For two different lines ($i\neq i'$), the $C_ {\ell,\nu\nu'ii'}$'s are only nonzero when the two channels probe the two emission lines $i$ and $i'$ from the same redshift. 
Figure~\ref{F:signal_model_Cl} 
presents an example angular-power-spectrum matrix $\mathbf{C}^{\rm line}_{\ell}$, where the line signal model is detailed in Section~\ref{S:line_modeling}.

Under the assumption that the noise is white noise without cross-channel correlation, $\mathbf{C}^{n}_{\ell}$ is an $\ell$-independent diagonal matrix:
\begin{equation}\label{E:Cln}
C_{\ell,\nu\nu'}^n = \sigma_n^2(\nu)\Omega_{\rm pix}\delta^K_{\nu,\nu'},
\end{equation}
where $\delta^K$ is the Kronecker delta.

In practice, the data usually exhibit correlated noise from the instrument and/or foreground residuals. In Section~\ref{S:corr_noise}, we present the results of applying our algorithm in the presence of such correlated noise.

Finally, there are stochastic fluctuations in the real data, such that the observed power spectrum of the data at a multipole bin $\ell$, $\mathbf{C}^d_{\ell}$, is a random sample from a Wishart distribution with a scale matrix given by $\mathbf{C}_{\ell}$ (Equation~\ref{E:total_Cl}) and the degree of freedom $n_\ell$, where $n_\ell$ is the number of $\ell$ modes in the binned angular power spectrum, which depends on the bin width and the survey angular size, detailed in Section~\ref{S:survey_setup}.

\subsubsection{Caveats of our Power Spectrum Model}\label{S:Limber_RSD_caveats}
We employ the Limber approximation in this work. However, we note that there are a few caveats associated with this simplification. First, the accuracy of the Limber approximation depends on the comoving width of the window function $W_{i\nu}$. For narrow widths, the LOS modes may contribute to a non-negligible level. Furthermore, if there are lines with close rest-frame frequencies, such as the $[$\ion{O}{3}$]$ and H$\beta$ lines in the case we considered (see Section~\ref{S:survey_setup}), additional correlation will occur between the two lines due to the correlation between close LOS distances, even if the two lines do not fall in the same spectral channel. While this LOS correlation adds additional information for the inference, this is being ignored in our current implementation using the Limber approximation. All these effects can be properly taken into account by using the exact expression in Equation~\ref{E:Cl_line_bessel} instead of relying on the Limber approximation, albeit at the cost of computing triple integrations. While these are important considerations in practice, for the purpose of demonstrating our technique, we defer more detailed investigations to future work.

Here, we ignore the RSD effect in our model. The RSD effect would introduce additional terms to the angular power spectrum in Equation~\ref{E:Cl_line_bessel}, helping to break the degeneracy between $b_i(\chi)$ and $M_{0,i}(\chi)$ in the window function. By using the Limber approximation in Equation~\ref{E:Cl_line}, our formalism only accounts for the transverse modes, which are not impacted by the RSD effect. Therefore, without the inclusion of RSD, our focus is on constraining the quantity $b_i(\chi)M_{0,i}(\chi)$ from the data rather than the two terms individually. More detailed investigations considering the RSD effect are left for future work.

\section{Survey Setup}\label{S:survey_setup}
\begin{figure}[ht!]
\begin{center}
\includegraphics[width=\linewidth]{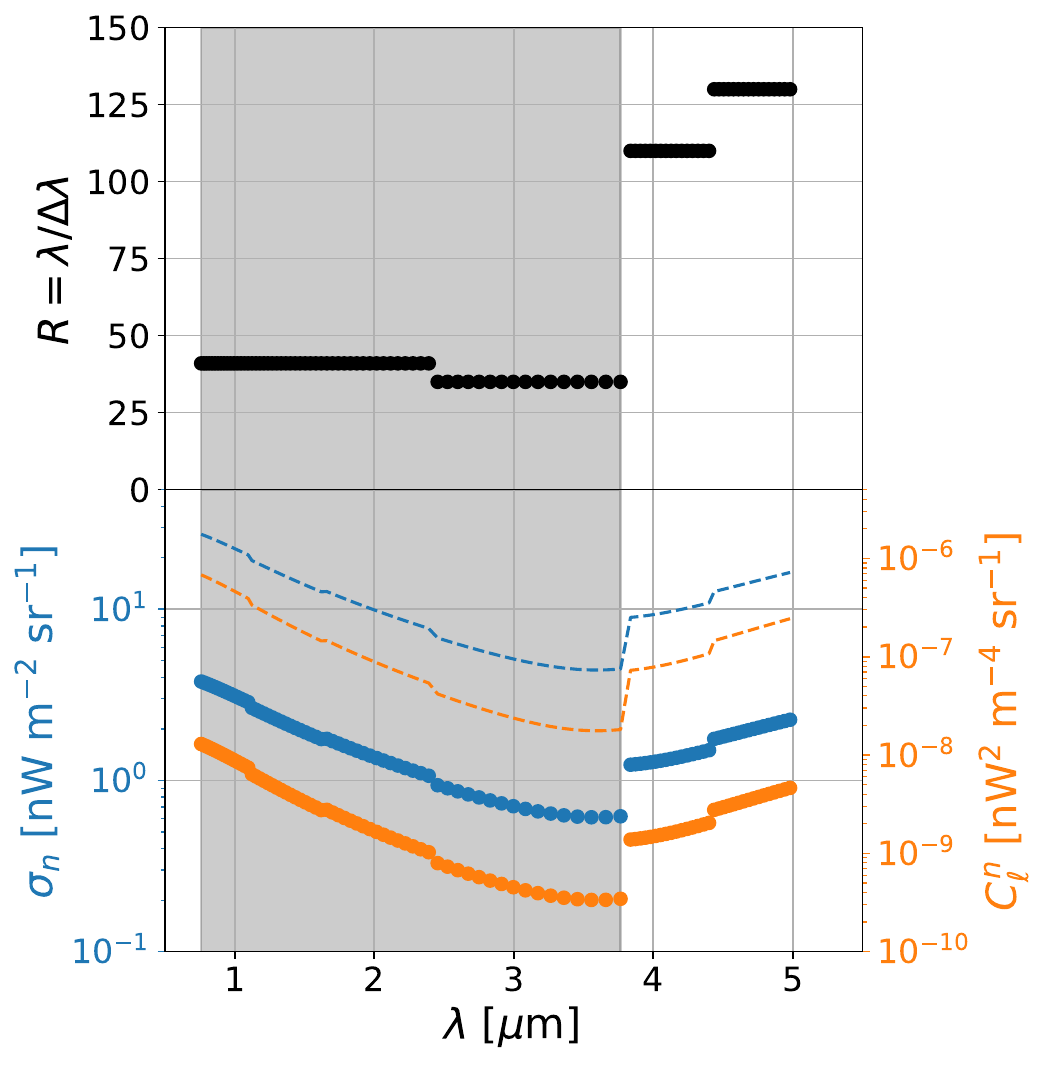}
\caption{\label{F:spherex_R_noise} Top: SPHEREx spectral resolution of each channel. Bottom: SPHEREx surface brightness sensitivity per spectral channel in a $6''.2$ sky pixel (blue) and the corresponding instrument noise power spectrum in the deep fields (Equation~\ref{E:Cln}; orange). The solid points and dashed lines represent the SPHEREx all-sky and deep-field sensitivity, respectively. We use the 96-channel configuration from the SPHEREx public products. The gray shaded region marks the 64 channels considered in this work ($0.75$--$3.82$ $\mu$m).}
\end{center}
\end{figure}

We demonstrate our technique with a survey setup similar to the deep-field survey of the SPHEREx mission\footnote{\url{http://spherex.caltech.edu}} \citep{2014arXiv1412.4872D, 2016arXiv160607039D, 2018arXiv180505489D}. 
SPHEREx is the next NASA Medium Class Explorer mission scheduled to launch in early 2025. SPHEREx will carry out the first all-sky near-infrared spectro-imaging survey from 0.75 to 5 $\mu$m with a pixel size of $6''.2$ through four consecutive surveys over the nominal 2 yr mission. SPHEREx consists of six H2RG detector arrays spanning six broad bands in the near-infrared, with low-resolution spectroscopy conducted by linear variable filters \citep{2018SPIE10698E..1UK}. Each band contains 17 channels with different spectral resolutions: $R=41$ for bands 1--3 at wavelengths between 0.75 and 2.42 $\mu$m, (51 spectral channels), $R=35$ for band 4 between 2.42 and 3.82 $\mu$m (17 spectral channels), $R=110$ for band 5 between 3.82 and 4.42 $\mu$m (17 spectral channels), and $R=130$ for band 6 between 4.42 and 5.00 $\mu$m (17 spectral channels). SPHEREx will scan the north and south ecliptic poles with a much higher cadence, due to its scanning strategy. Consequently, SPHEREx will produce two deep field mosaic maps of $\sim 100$ deg$^2$ each, with the noise rms $\sim 50$ times lower than its all-sky survey (Figure~\ref{F:spherex_R_noise}).

\begin{deluxetable}{c|c|c|c|c}[h]
\tablenum{1}
\tablecaption{\label{T:spherex_lines} Spectral lines modeled in this work}
\tablewidth{0pt}
\tablehead{
\colhead{Line}  & \colhead{$\lambda_{\rm rf}$} & \colhead{SPHEREx Coverage} & \colhead{$r_i$} & \colhead{$A_i$} 
}
\startdata
H$\alpha$ & $0.6563$ & $0.14 < z < 4.82\,(6.62)$ & $1.27$ & $1.0$\\
$[$\ion{O}{3}$]$ & $0.5007$ & $0.50 < z < 6.63\,(8.99)$ & $1.32$ & $1.32$\\
H$\beta$ & $0.4861$ & $0.54 < z < 6.68\,(9.29)$ & $0.44$ & $1.38$\\
$[$\ion{O}{2}$]$ & $0.3727$ & $1.01 < z < 9.25\,(12.4)$ & $0.71$ & $0.62$\\
\hline\hline
\enddata
\tablecomments{$\lambda_{\rm rf}$: rest-frame wavelength ($\mu$m); The maximum redshift, with and without parentheses, correspond to the first four bands and the full SPHEREx coverage with $\lambda=3.82$ and $5$ $\mu$m, respectively; $r_i$: line luminosity and the SFR ratio $L_i$/SFR ($10^{41}$~erg~s$^{-1}$~$M_\odot^{-1}$~yr). $A_i$: dust extinction factor (mag).}
\end{deluxetable}

Here, we consider a survey setup similar to SPHEREx deep fields totalling $200$ deg$^2$ ($100$ deg$^2$ in each ecliptic pole) in this study, corresponding to a sky fraction of $f_{\rm sky}=0.48\%$\footnote{SPHEREx will also provide all-sky coverage with shallower depth, which can potentially be used for LIM analysis. However, the lower redundancy over all sky will make reliable all-sky maps harder to construct. Therefore, we focus on the deep field in this study.}. We use the first four bands of SPHEREx spanning 0.75 to 3.82 $\mu$m and assume nonoverlapping top-hat filters equally spaced in the logarithmic frequency. The last two bands are not included since they only probe the very high redshift emission from the four lines we consider in the SPHEREx spectral coverage (see Table~\ref{T:spherex_lines}). We consider an angular resolution of $6''.2$, and the surface brightness sensitivity in each channel given by the public products\footnote{Data downloaded from: \url{https://github.com/SPHEREx/Public-products/blob/master/Surface_Brightness_v28_base_cbe.txt}}. The SPHEREx public products are based on the previous design, which has 16 instead of 17 spectral channels in each band, and thus we also use the same configuration of 16 channels per band in this work, which gives 64 channels (four bands) in total.

The top panel of Figure~\ref{F:spherex_R_noise} shows the SPHEREx spectral resolution as a function of wavelength (spectral channel), and the bottom panel shows the expected noise rms per spectral channel per pixel in SPHEREx and the corresponding noise power spectrum given by Equation~\ref{E:Cln}.

We consider the following four lines within the SPHEREx spectral coverage: H$\alpha$, $[$\ion{O}{3}$]$, H$\beta$, and $[$\ion{O}{2}$]$. Table~\ref{T:spherex_lines} summarizes their rest-frame wavelengths and the redshift ranges that SPHEREx can probe. For the purpose of demonstrating our technique, we only focus on these four lines in this work. We note that there are more spectral lines within the SPHEREx spectral range---such as Ly$\alpha$, Paschen-$\alpha$, $[$\ion{N}{2}$]$, and $[$\ion{S}{2}$]$---and the line flux of some of them may be comparable to the four lines considered here \citep{2023arXiv231204636F}. Therefore, in practice, the analysis for SPHEREx should account for all the prominent lines for a more realistic modeling.

We consider the line emission from sources within the redshift range $0.7 < z < 6$. Removing detected local point sources at lower redshift helps improve the sensitivity in probing the diffuse line emission from fainter sources. We estimate that below our chosen redshift lower limit $z_{\rm min} = 0.7$, we can reliably detect and constrain the redshifts of the majority of galaxies with SPHEREx, thus allowing them to be masked before calculating the power spectrum (see Appendix~\ref{A:ellmax} for details). The choice of the maximum redshift $z_{\rm max}$ does not affect our results; as shown in Figure~\ref{F:bnuInu_z_constraints}, we have no sensitivity on the line signal at $z\gtrsim4$.

We choose the multipole mode range of $50<\ell<350$ in our analysis. The minimum $\ell$ mode ($\ell_{\rm min}=50$) corresponds to SPHEREx’s field of view of $3.5^\circ$ (on the smaller side), as fluctuations larger than the field size will be partially suppressed due to zodiacal light filtering in processing individual exposures and we do not model this effect. The choice of the maximum $\ell$ mode ($\ell_{\rm max}=350$) is made to restrict our analysis to linear clustering scale (see Appendix~\ref{A:ellmax} for details). 

We use eight $\ell$ bins within the range of $50<\ell<350$. The bins are selected to contain approximately the same number of modes in each bin. For a given $\ell$ bin spanning $\ell \in [\ell^\alpha_{\rm min}, \ell^\alpha_{\rm max})$, the number of multipole modes $n^\alpha_{\ell}$ is given by
\begin{equation}\label{E:nell}
n^\alpha_{\ell} = f_{\rm sky}\left[(\ell^\alpha_{\rm max})^2 - (\ell^\alpha_{\rm min})^2\right],
\end{equation}
where $f_{\rm sky}=0.48\%$ is the fraction of sky area in the $200$ deg$^2$ field. In reality, some pixels with bright foreground contamination will be masked, resulting in a reduction of the effective number of modes. We ignore this effect in our analysis.

\section{Line Signal Modeling}\label{S:line_modeling}
Our model for line emission follows the prescription from \citet{2017ApJ...835..273G}, which is built on an empirical star-formation rate (SFR) and the line luminosity relation. We use the SFR density (SFRD) constraints to model the luminosity density for each line.

We assume a linear relation between the line luminosity $L_i$ and the SFR:
\begin{equation}
\left[\frac{L_i}{{\rm erg\,s}^{-1}}\right] = r_i \left[\frac{{\rm SFR}}{M_\odot\,{\rm yr}^{-1}}\right],
\end{equation}
and use the $L_i$--SFR relations from \citet{1998ARA&A..36..189K} and \citet{2007ApJ...657..738L} for H$\alpha$, $[$\ion{O}{2}$]$, and $[$\ion{O}{3}$]$. For H$\beta$, we assume a fixed line ratio of $L_{H\beta}/L_{H\alpha}=0.35$ \citep{2006agna.book.....O}, which has been validated to have good agreement with simulations and observations by \citet{2017ApJ...835..273G}. The line luminosity--SFR ratio ($r_i$) for each line is summarized in Table~\ref{T:spherex_lines}. Following \citet{2017ApJ...835..273G}, we adopt the same dust extinction factors, also listed in Table~\ref{T:spherex_lines}. 

Despite this model adopting a simple linear scaling for the $L_i$--SFR relation, \citet{2017ApJ...835..273G} have validated that the resulting line intensity is in agreement with another model based on simulations as well as the observational constraints from integrating the observed line luminosity functions. We also note that there are scatters in the $L_i$--SFR relation in reality, which will boost the power spectrum amplitude \citep{2019ApJ...887..142S}. For simplicity, we ignore the effect of this scatter in this work.

For the SFRD, we use the analytical fitting formula from \citet{2014ARA&A..52..415M}:
\begin{equation}
\frac{{\rm SFRD}(z)}{M_\odot\,{\rm yr}^{-1}\,{\rm Mpc}^{-3}} = 0.015\, \frac{(1+z)^{2.7}}{1+\left[(1+z)/2.9\right]^{5.6}}.
\end{equation}
The luminosity density of the line is then given by a linear scaling of the SFRD:
\begin{equation}
M_{0,i}(z) = \frac{dL_i(z)}{dV} = r_i \cdot {\rm SFRD}(z).
\end{equation}

We model the luminosity-weighted bias $b_i$ of the lines with the halo-mass-weighted bias, under the assumption that the line luminosity is proportional to the halo mass:
\begin{equation}
b_i(z) = \frac{\int dM \frac{dn}{dM}(M,z)b_h(M,z)}{\int dM \frac{dn}{dM}(M,z)},
\end{equation}
where $\frac{dn}{dM}$ is the halo mass function \citep{1999MNRAS.308..119S} and $b_h$ is the halo bias \citep{2001MNRAS.323....1S}. Our bias prescription assumes the linear relation between the line luminosity and the halo mass. We assume the same $b_i(z)$ for all spectral lines. Note that although here we build models for $M_{0,i}(z)$ and $b_i(z)$ separately, our method can only constrain $M_{i}(z)$, the product of these two terms.

\begin{figure}[ht!]
\begin{center}
\includegraphics[width=\linewidth]{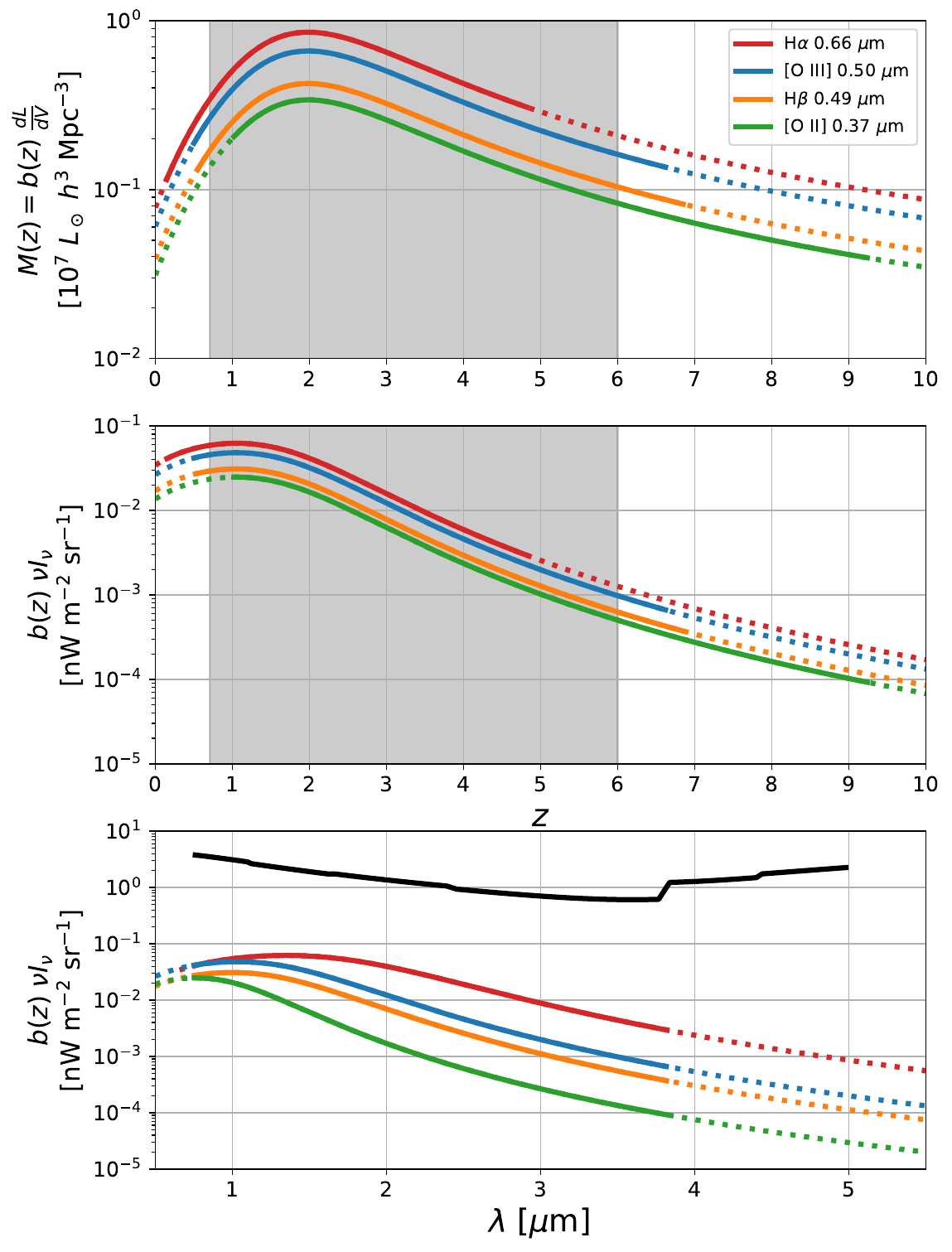}
\caption{\label{F:signal_model} Top: our model of the bias-weighted luminosity density $M_i(z)$ as a function of redshift. Middle: the bias-weighted line intensity as a function of redshift. The gray shaded regions in the top and middle panels denote the range of redshift considered in this work ($0.7<z<6$). Bottom: the bias-weighted line intensity as a function of observed wavelength. The solid parts in all three panels denote the redshift/wavelength ranges that can be probed by the spectral range considered in this work (0.75--3.82 $\mu$m). The SPHEREx noise level $\sigma_n$ is also shown at the bottom panel for comparison (black). We note that while noise fluctuations overwhelm the line intensity, the large-scale clustering power of the lines is not suppressed by the noise power, as shown in Figure~\ref{F:signal_model_Cl}.}
\end{center}
\end{figure}

Figure~\ref{F:signal_model} shows our model of the bias-weighted luminosity density $M_i(z)$ and the bias-weighted intensity $b_i(z)\nu I_{\nu}^i(z)$ of each line (see Equation~\ref{E:nuInu_line} for the conversion from luminosity density to intensity). Since our line modeling is a linear scaling of the SFRD, $M_i(z)$ follows the same redshift dependence as the SFRD, which exhibits a peak at $z\sim 2$. The $M_i(z)$ from different lines are linearly proportional to each other in our model, since we assume the same line bias and a linear scaling from SFRD for all lines.

While this simple model is not able to capture the complex line emission mechanisms in reality, it is sufficient for our purpose of demonstrating the algorithm in this study. Furthermore, as detailed in Section~\ref{S:parameterization}, we introduce a flexible parameterization to fit for any redshift dependence of the $M_i(z)$ function, which is not restricted to a certain functional form of $M_i(z)$.

\begin{figure}[ht!]
\begin{center}
\includegraphics[width=\linewidth]{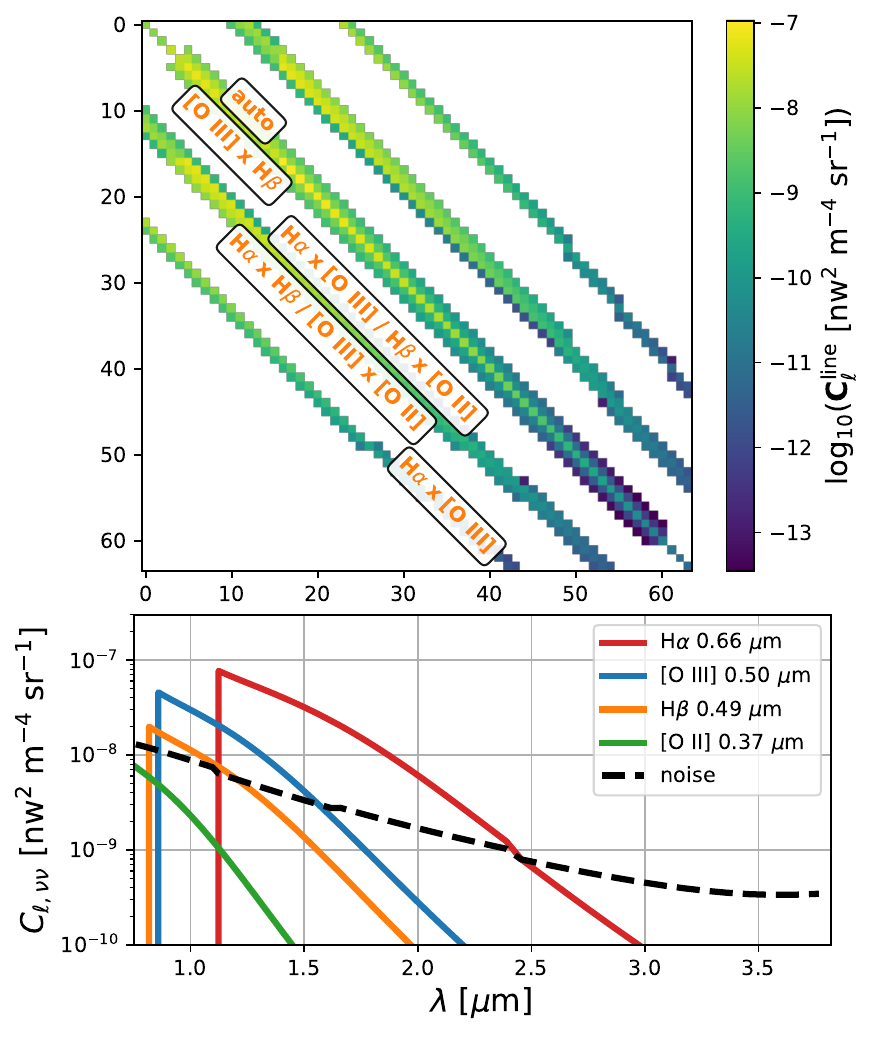}
\caption{\label{F:signal_model_Cl} Top: the line power spectrum matrix $\mathbf{C}_\ell^{\rm line}$ (Equation~\ref{E:Cl_line_matrix}) in the lowest-$\ell$ mode centered at $\ell=91$. The signals from different pairs of lines are labeled in the figure. Bottom: the auto power spectrum of each line (colored lines) compared to the SPHEREx noise power spectrum (black dashed line). The short-wavelength cutoff corresponds to the minimum redshift $z_{\rm min}=0.7$, where we assume no line signal below this redshift, as galaxies can effectively be masked given the SPHEREx deep-field depth (Appendix~\ref{A:ellmax}).}
\end{center}
\end{figure}

The top panel of Figure~\ref{F:signal_model_Cl} presents the line power spectrum matrix $\mathbf{C}_{\ell}^{\mathrm{line}}$ from our model at the lowest-$\ell$ bin centered at $\ell=91$. The diagonal power is from the autocorrelation of each line at the same redshift/frequency. The ``broadened'' band along the diagonal line is the cross-power of the closely paired $[$\ion{O}{3}$]$ and H$\beta$ lines. Other off-diagonal correlations arise from different combinations of cross-power between pairs of lines, as marked in Figure~\ref{F:signal_model_Cl}. Coincidentally, the cross-power of H$\alpha$ and the $[$\ion{O}{3}$]$-H$\beta$ pair falls at overlapping frequency channels with $[$\ion{O}{2}$]$ and the $[$\ion{O}{3}$]$-H$\beta$ pair.

The bottom panel of Figure~\ref{F:signal_model_Cl} shows the autospectrum of each line compared to the SPHEREx noise power spectrum. Although the SPHEREx noise overwhelms the mean intensity of all the line signals (Figure~\ref{F:signal_model}), the line emission traces the underlying density field, exhibiting large-scale clustering that boosts the line power spectrum above the noise fluctuations on large scales (Figure~\ref{F:signal_model_Cl}). Moreover, we emphasize that in the power spectrum space, from which we extract information, the noise is present only in the diagonal elements. The unique off-diagonal features in the frequency--frequency correlation space are unaffected by noise fluctuations.

\section{Algorithm}\label{S:algorithm}
This section describes our algorithm for constraining the line signals from the LIM data. Our method infers the bias-weighted luminosity density ($M_i(z)$) for each line from the auto- and cross-frequency power spectra ($C_{\ell,\nu\nu'}$'s). We first introduce our parameterization for $M_i(z)$ (Section~\ref{S:linear_basis}), then we describe our Bayesian inference framework (Section~\ref{S:bayesian_framework}) and the algorithm for inferring the parameter constraints (Section~\ref{S:parameter_inference}).

\subsection{Parameterization}\label{S:parameterization}
Our goal is to infer the function $M_i(z)$ for each line from the LIM data. This is achieved by first parameterizing $M_i(z)$, and then fitting the defined parameters to constrain the $M_i(z)$ functions. $M_i(z)$ can be characterized with a small number of parameters as $M_i(z)$ is expected to be a smooth function with redshift. One approach is to use a parametric functional form to define a smooth curve for $M_i(z)$. However, here we instead choose a series of linear basis models for $M_i(z)$. As detailed in Section~\ref{S:linear_basis}, this parameterization allows us to precompute the basis power spectra , $\hat{C}_{\ell,\nu\nu}$'s, to significantly reduce the computational cost during the inference stage. To ensure flexibility in capturing any possible redshift dependence of the signal, we employ a basis function set that forms a piecewise linear function for $M_i(z)$ (Section~\ref{S:basis_functions}). The piecewise linear function can approximate any continuous functions, and thus we are not restricted to any prior assumptions about the shape of $M_i(z)$ functions.

\subsubsection{Linear Basis Decomposition}\label{S:linear_basis}
To parametrize $M_i(z)$, we decompose it with a linear combination of basis functions $\{\hat{M}_j(z)\}$:
\begin{equation}\label{E:M_linear_decomposition}
M_i(z) = \sum_{j=1}^{N_m}c_{ij}\hat{M}_j(z),
\end{equation}
and we fit for the coefficients $\{c_{ij}\}$ to constrain $M_i(z)$ for each line. With this decomposition, we can also express the power spectrum in terms of the linear combinations of the basis functions:
\begin{equation}
C_{\ell,\nu\nu'ii'jj'} = \sum_{j=1}^{N_m}\sum_{j'=1}^{N_m}c_{ij}c_{i'j'}\hat{C}_{\ell,\nu\nu'ii'jj'},
\end{equation}
where
\begin{equation}\label{E:Cl_nunu1ii1_basis}
\begin{split}
\hat{C}_{\ell,\nu\nu'ii'jj'}
&= \frac{\nu}{\Delta\nu}\frac{\nu'}{\Delta\nu'}\frac{\Delta\chi_{ii'\nu\nu'}}{\chi^2_{ii'\nu\nu'}}A^2(\chi_{ii'\nu\nu'})\\
&\cdot \hat{M}_j(\chi_{ii'\nu\nu'})\hat{M}_{j'}(\chi_{ii'\nu\nu'})P(\frac{\ell+\frac{1}{2}}{\chi_{ii'\nu\nu'}}).
\end{split}
\end{equation}

The total power spectrum from all the spectral lines can then be written as
\begin{equation}\label{E:Cl_line_matrix}
\mathbf{C}_{\ell}^{\rm line} = \sum_{i=1}^{N_{\rm line}}\sum_{i'=1}^{N_{\rm line}}\sum_{j=1}^{N_m}\sum_{j'=1}^{N_m} c_{ij}c_{i'j'}\hat{\mathbf{C}}_{\ell,ii'jj'}.
\end{equation}
In the parameter inference stage (Section~\ref{S:parameter_inference}), we also need the derivatives of the power spectrum with respect to the parameters. This can also be expressed as the linear combination of the basis components:
\begin{equation}\label{E:dCl_dcij}
\frac{\partial \mathbf{C}_{\ell}}{\partial c_{ij}}=\frac{\partial \mathbf{C}_{\ell}^{\rm line}}{\partial c_{ij}} = 2\sum_{i'=1}^{N_{\rm line}}\sum_{j'=1}^{N_m}c_{i'j'}\hat{\mathbf{C}}_{\ell,ii'jj'}.
\end{equation}

With this linear basis decomposition, the basis power spectra $\hat{\mathbf{C}}_{\ell,ii'jj'}$ can be precomputed to greatly speed up the inference process.

\subsubsection{Basis Functions}\label{S:basis_functions}
We use a piecewise linear function to describe the bias-weighted luminosity density $M_i(z)$. The piecewise linear function can flexibly approximate any continuous function with a sufficiently fine segmentation, and it does not depend on any underlying assumption about the shape of the function being fitted. Since $M_i(z)$ follows the global redshift dependence of the large-scale bias and the SFRD, it is expected to be a smooth function of redshift. Thus, only a few segments of the piecewise linear function are sufficient to approximate the $M_i(z)$ functions.

\begin{figure}[ht!]
\begin{center}
\includegraphics[width=\linewidth]{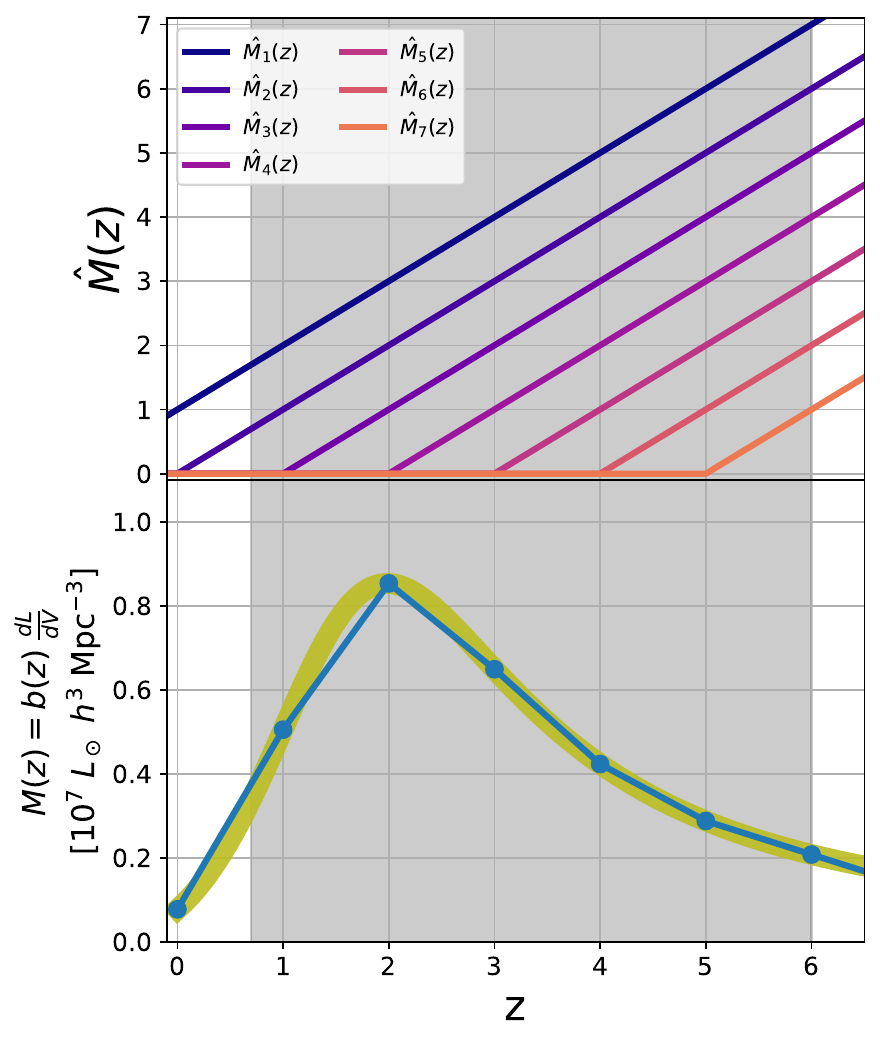}
\caption{\label{F:M_basis} Top: the seven ReLU basis functions $\{\hat{M}_j(z)\}$ for decomposing the bias-weighted luminosity density ($M_i(z)$) for each line. Bottom: our fiducial input model $M_i(z)$ (the blue solid line with dots). This is a piecewise linear function with anchoring redshifts at $z=0,1,...,6$ (blue dots) that fit to the modeled $M_i(z)$ (yellow curve; the same as the top panel of Figure~\ref{F:signal_model}). This piecewise linear function can be produced by the linear combinations of the seven basis functions $\{\hat{M}_j(z)\}$ shown in the top panel. Here, we show the H$\alpha$ line as an example. The fiducial model for other lines is set with the same process. The gray shaded region denotes our redshift range of $0.7 < z < 6$.}
\end{center}
\end{figure}

The piecewise linear function can be expressed in terms of linear combinations of a series of rectified linear unit (ReLU) functions. Thus, we define our basis functions $\{\hat{M}_j(z)\}$ as ReLU functions with anchoring redshifts $z_j$'s:
\begin{equation}\label{E:M_basis_relu}
\begin{split}
\hat{M}_j(z) &=
\begin{cases}
z - z_j
& \quad\text{if } z_j < z\\
0 & \quad\text{otherwise}.
\end{cases}
\end{split}
\end{equation}
We choose $\{z_j\} = \{-1,0,...,5\}$, which gives a total number of basis functions $N_m=7$. The basis functions $\hat{M}_j(z)$ are shown in the top panel of Figure~\ref{F:M_basis}.

The linear combination of this basis set spans the piecewise linear functions anchored at $z=z_j+1$ (i.e., $z=0,1,...,6$). This provides the flexibility to approximate any underlying $M_i(z)$ functions, and we can also easily increase the accuracy at any specific redshift range by adding more ReLU basis functions with $z_j$ around the desired redshifts. The optimal number of redshift anchoring points and their positions depend on the line signals and noise of particular surveys. Therefore, we leave this further investigation to future work.

While $\{c_{ij}\}$ is the native parameter set in our formalism, the underlying line signal is best described by $\{m_{ij}\}$, defined as
\begin{equation}
m_{ij} = M_i(z_j+1),
\end{equation}
which is the $M_i$ value at the anchoring redshifts $z=z_j+1$. There is a simple linear transformation between $\{m_{ij}\}$ and $\{c_{ij}\}$, detailed in Appendix~\ref{A:parameter_transformation}. This transformation relation enables us not only to convert values between $\{m_{ij}\}$ and $\{c_{ij}\}$, but also to propagate the parameter constraints, which are determined by the Jacobian of this transformation.

\subsubsection{Fiducial Parameters}\label{S:fiducial_params}
Our fiducial input parameters $\{c_{ij}\}$ are set by matching their corresponding $\{m_{ij}\}$ to the modeled $M_i(z)$ (the top panel of Figure~\ref{F:signal_model}) at the seven anchoring redshifts ($z=0,1,...,6$), as demonstrated in the bottom panel of Figure~\ref{F:M_basis}.

In reality, our piecewise linear model serves as an approximate representation of the true signal. Nonetheless, we employ this piecewise linear approximation as the fiducial input, providing a set of ground-truth parameters for evaluating our algorithm's performance.

We emphasize that while our fiducial input assumes the same shape for $M_i(z)$ across all four lines, our algorithm fits each line separately. This allows us to reconstruct the redshift evolution of $M_i(z)$ for each line independently. Further demonstration of this capability is provided in Section~\ref{S:model_misspecification}, where different $M_i(z)$ functions are used to generate the mock signal for each line, and we validate that our algorithm can robustly extract the inputs, even when the input $M_i(z)$'s are smooth curves that are not able to be perfectly described by our piecewise linear parameterization. The example in Section~\ref{S:model_misspecification} also assumes very different shapes of $M_i(z)$ for each line, to demonstrate the algorithm's robustness against model variations.

\subsection{Bayesian Framework}\label{S:bayesian_framework}
Our parameter inference method follows the framework presented in C23. We constrain the parameter set $\mathbf{\Theta}$ from the data power spectra $\{\mathbf{C}^d_\ell\}$ in $N_\ell$ multipole bins using a Bayesian framework. The posterior probability distribution $p\left ( \mathbf{\Theta}|\{\mathbf{C}^d_\ell\}\right )$ is given by
\begin{equation}
p\left ( \mathbf{\Theta}|\{\mathbf{C}^d_\ell\}\right ) \propto \mathcal{L}\left ( \{\mathbf{C}^d_\ell\}|\mathbf{\Theta}\right ) \pi\left ( \mathbf{\Theta}\right ),
\end{equation}
where $\mathcal{L}$ and $\pi$ are the likelihood and prior, respectively. 

Here, our parameter set $\mathbf{\Theta}$ consists of the coefficients of the basis components for $M_i(z)$ for each line, i.e., $\{c_{ij}\}$. For the prior, we only enforce the positivity condition on $M_i(z)$, which is equivalent to requiring all $m_{ij}$ values to be positive in our piecewise linear model. Here, $m_{ij}$ represents the $M_i(z)$ function at the anchoring redshifts in our piecewise linear model, thus enforcing the positivity of $m_{ij}$ guarantees that $M_i(z)$ is positive across all redshifts. We implement this positivity constraint through a logarithmic transformation on $m_{ij}$, defining a new set of parameters $\{\theta_{ij}\}$, where $\theta_{ij}=\log m_{ij}$ (see Appendix~\ref{A:parameter_transformation} for details). Our objective is to determine the maximum a posteriori solution for $\{\theta_{ij}\}$. By doing so, the positivity constraint on $m_{ij}$'s will be automatically satisfied. We set flat priors on $\{\theta_{ij}\}$, which effectively give the logarithmic priors on $\{m_{ij}\}$.\footnote{In practice, one can incorporate external information, such as line luminosity function constraints, into priors. The choice of different priors may have a non-negligible impact on inference \citep{2018A&A...617A..96M}. We leave these considerations for future work.}

As we consider the two-point information in this work, the likelihood of the full LIM dataset, i.e. the voxel intensities, can be described as a Gaussian distribution, and the cross angular power spectrum $C_{\ell,\nu\nu'}$'s represent the covariance matrices of the Gaussian likelihood on the voxel intensity maps in the spherical harmonic space.
As each multipole mode is independent, the log-likelihood function is the sum of normal distributions $\mathcal{N}$ for each $\ell$ bin:
\begin{equation}\label{E:likelihood}
\begin{split}
{\rm log} \,\mathcal{L}\left ( \{\mathbf{C}^d_\ell\}|\mathbf{\Theta}\right )&=
-\frac{1}{2}\sum_\ell n_\ell\, {\rm log}\,\mathcal{N}\left ( \mathbf{C}^d_\ell,\mathbf{C}_\ell\left ( \mathbf{\Theta} \right ) \right )\\
&=-\frac{1}{2}\sum_\ell n_\ell\left [\right. {\rm Tr}\left ( \mathbf{C}^d_\ell \mathbf{C}_\ell^{-1}\left ( \mathbf{\Theta} \right ) \right )\\
&+{\rm log}\left | \mathbf{C}_\ell\left ( \mathbf{\Theta} \right ) \right | +N_\nu {\rm log} \left(2\pi\right)\left.\right ],
\end{split}
\end{equation}
where $n_\ell$ is the number of modes in each $\ell$ bin (Equation~\ref{E:nell}), $\mathbf{C}_\ell\left ( \mathbf{\Theta} \right )$ is the modeled power spectrum given the parameter set $\mathbf{\Theta}$.

We note that our likelihood models the voxel intensity, the native data product from LIM, as a Gaussian distribution with covariances given by the $C_{\ell,\nu\nu'}$'s. This captures the full two-point information in the field, which is a lossless representation on large scales, where the underlying signal is expected to be fully characterized by a Gaussian distribution.

\subsection{Parameter Inference}\label{S:parameter_inference}
With the observed angular power spectra from the LIM data $\{\mathbf{C}_\ell^d\}$, we conduct parameter inference within the Bayesian framework outlined in Section~\ref{S:bayesian_framework}. The inference process is similar to our prior work in C23, involving two steps: first, we employ the Newton--Raphson method to identify the parameter set corresponding to the maximum likelihood; then, we estimate the parameter constraints using the Fisher matrix derived from the maximum likelihood found with the Newton--Raphson method.

\subsubsection{Newton--Raphson Method}\label{S:newtons}
The Newton--Raphson method is an iterative approach to find the maximum/minimum of a function. In this context, our objective is to find the parameter set $\Theta_{\rm max}$ that maximizes the log-likelihood, given the angular power spectra from the data, $\{\mathbf{C}_\ell^d\}$:
\begin{equation}
\Theta_{\rm max} = \max_{\Theta} {\rm log}\,\mathcal{L}\left ( \{\mathbf{C}^d_\ell\}|\mathbf{\Theta}\right ).
\end{equation}

The Newton--Raphson algorithm iteratively updates the current parameter set from $\mathbf{\Theta}_t$ to $\mathbf{\Theta}_{t+1}$ through the equation:
\begin{equation}
\mathbf{\Theta}_{t+1} = \mathbf{\Theta}_t - \eta\, \mathbf{H}^{-1} \mathbf{g}
\end{equation}
where $\mathbf{g}=\nabla\,{\rm log}\,\mathcal{L}$ and $\mathbf{H}=\nabla\,({\rm log}\,\mathcal{L})\,\nabla^T$ represent the gradient and Hessian matrix of the log-likelihood, respectively. For the complete expressions and detailed derivations, see Appendix E of C23. The parameter $\eta$ serves as the learning rate, determining the step size of the update. In each iteration of the Newton--Raphson update, we initiate with $\eta=2$ and subsequently verify whether the new proposed parameter set yields a higher log-likelihood value. If not, we reduce the value of $\eta$ by half until the condition is satisfied.

The Newton--Raphson optimization is performed with the logarithmically transformed parameter set $\{\theta_{ij}\}$. In this parameter space, the positivity condition for the prior is automatically satisfied, eliminating the need for additional prior constraints, such as a hard boundary for the disallowed parameter space.

Similar to the algorithm in C23, we utilize an approximated Hessian provided by Equation E33 in C23 to avoid computing the second derivatives of $C_\ell$ on parameters, thus expediting the optimization process. This approximation approaches the exact expression when $\mathbf{\Theta}_t$ is in proximity to $\mathbf{\Theta}_{\rm max}$, and we have confirmed the successful convergence of our algorithm using this approximation.

\subsubsection{Fisher Matrix}\label{S:fisher}
After determining $\mathbf{\Theta}_{\rm max}$ using the Newton--Raphson method, we estimate the parameter constraints with the Fisher matrix at $\mathbf{\Theta}_{\rm max}$. The Fisher matrix is given by
\begin{equation}\label{E:Fisher}
\begin{split}
\mathbf{F}_{\alpha\beta} &= -\left \langle \frac{\partial^2 {\rm log}\,\mathcal{L}}{\partial\theta_\alpha\partial\theta_\beta} \right \rangle\\
&=\frac{1}{2}\sum_\ell n_\ell {\rm Tr}\left ( \mathbf{C}_\ell^{-1}\frac{\partial\mathbf{C_\ell}}{\partial\theta_\alpha}\mathbf{C}_\ell^{-1}\frac{\partial\mathbf{C_\ell}}{\partial\theta_\beta} \right ),
\end{split}
\end{equation}
and the inverse of the Fisher matrix gives the covariance of the parameters:
\begin{equation}
{\rm Cov} (\theta_\alpha, \theta_\beta) = \mathbf{F}^{-1}_{\alpha\beta}.
\end{equation}

The Fisher matrix calculation requires the derivative of $\mathbf{C}_\ell$'s. The derivatives can also be expressed as linear combinations of the basis functions given by Equation~\ref{E:dCl_dcij}.

In Section~\ref{S:results}, we also quantify the constraints on $M_i(z)$ at any given redshift. From Equation~\ref{E:M_linear_decomposition}, we can obtain the covariance of $M_i(z)$ and $M_i'(z')$ for the two given lines $i$ and $i'$ at redshift $z$ and $z'$ by the expression
\begin{equation}\label{E:cov_Mz}
{\rm Cov}[M_i(z), M_{i'}(z')] = \sum_{j=1}^{N_m}\sum_{j'=1}^{N_m}{\rm Cov}[c_{ij}, c_{i'j'}]\hat{M}_j(z)\hat{M}_{j'}(z').
\end{equation}

\section{Results}\label{S:results}
\begin{figure*}[ht!]
\begin{center}
\includegraphics[width=\linewidth]{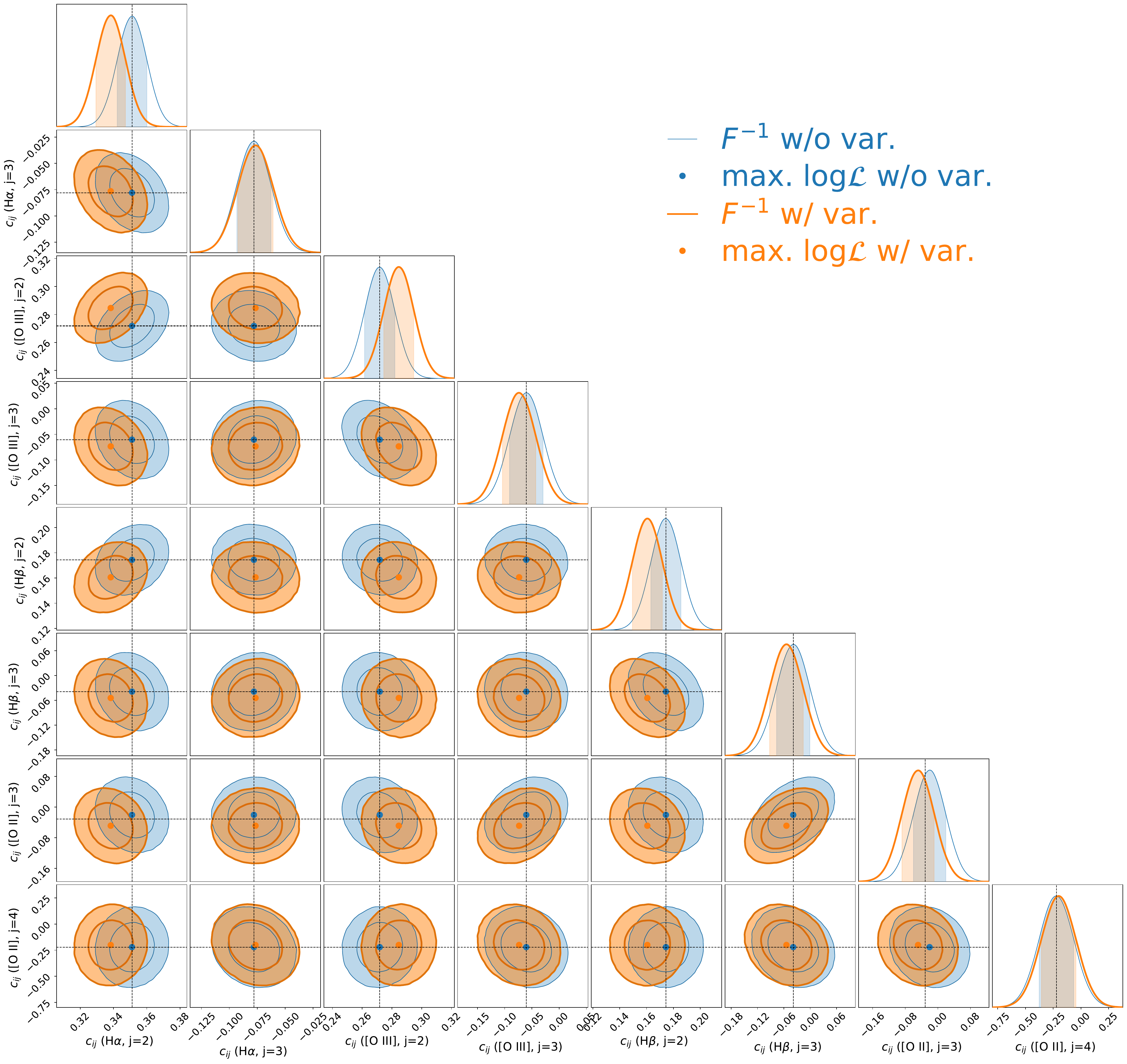}
\caption{\label{F:corner} Parameter constraints for our fiducial case. We display only the $c_{ij}$ constraints for the two basis components that are sensitive to the lowest-redshift line emission accessible by our survey setup for each line. The blue/orange colors represent the mock observed data power spectra $\{\mathbf{C}_\ell^d\}$ without/with sample variance fluctuations. The black dashed lines indicate the truth model input, while the blue/orange dots represent the maximum likelihood found by our Newton--Raphson algorithm. The contours show the 1 and 2$\sigma$ constraints derived from the Fisher matrix.}
\end{center}
\end{figure*}

Here, we present the results of inference using our fiducial model with the SPHEREx deep-field noise level. The mock observed power spectra $\{\mathbf{C}_\ell^d\}$ are produced by the fiducial parameter set described in Section~\ref{S:fiducial_params}. That is, we set $M_i(z)$'s to piecewise linear functions anchored at redshifts $z=0,1,...,6$, and the value of $M_i(z)$ for each line is fixed to an analytical model described in Section~\ref{S:line_modeling}. For each of the four lines, we fit for 7 linear coefficients $\{c_{ij}\}$ that define the $M_i(z)$ function. In summary, our data consist of the $64\times 64$ (64 spectral channels) symmetric matrices of $\{\mathbf{C}_\ell^d\}$ in eight $\ell$ bins, and we fit for 28 parameters (7 parameters for each of the 4 lines) in total.

With the input spectra $\{\mathbf{C}_\ell^d\}$, we first use the Newton--Raphson method to find the parameter set $\mathbf{\Theta}_{\rm max}$ that maximizes the log-likelihood function (Section~\ref{S:newtons}). The Newton--Raphson optimization in our case can efficiently converge within a few tens of steps. Then, we calculate the covariance on parameters using the Fisher matrix (Section~\ref{S:fisher}).

Figure~\ref{F:corner} displays the inference results on the fiducial model. We run the inference on cases with and without adding sample variance fluctuations to the data ($\{\mathbf{C}_\ell^d\}$), respectively. The case with sample variance fluctuations (orange contours) represents a realistic scenario, and the results show that our algorithm gives parameter constraints within about a 1$\sigma$ level of the truth, as expected. As a sanity check, we also run the case without sample variance fluctuations in the input data (blue contours). In this case, the likelihood function peaks at exactly the truth input values, verifying that our Newton--Raphson method can successfully locate the maximum a posteriori.

\begin{figure}[ht!]
\begin{center}
\includegraphics[width=\linewidth]{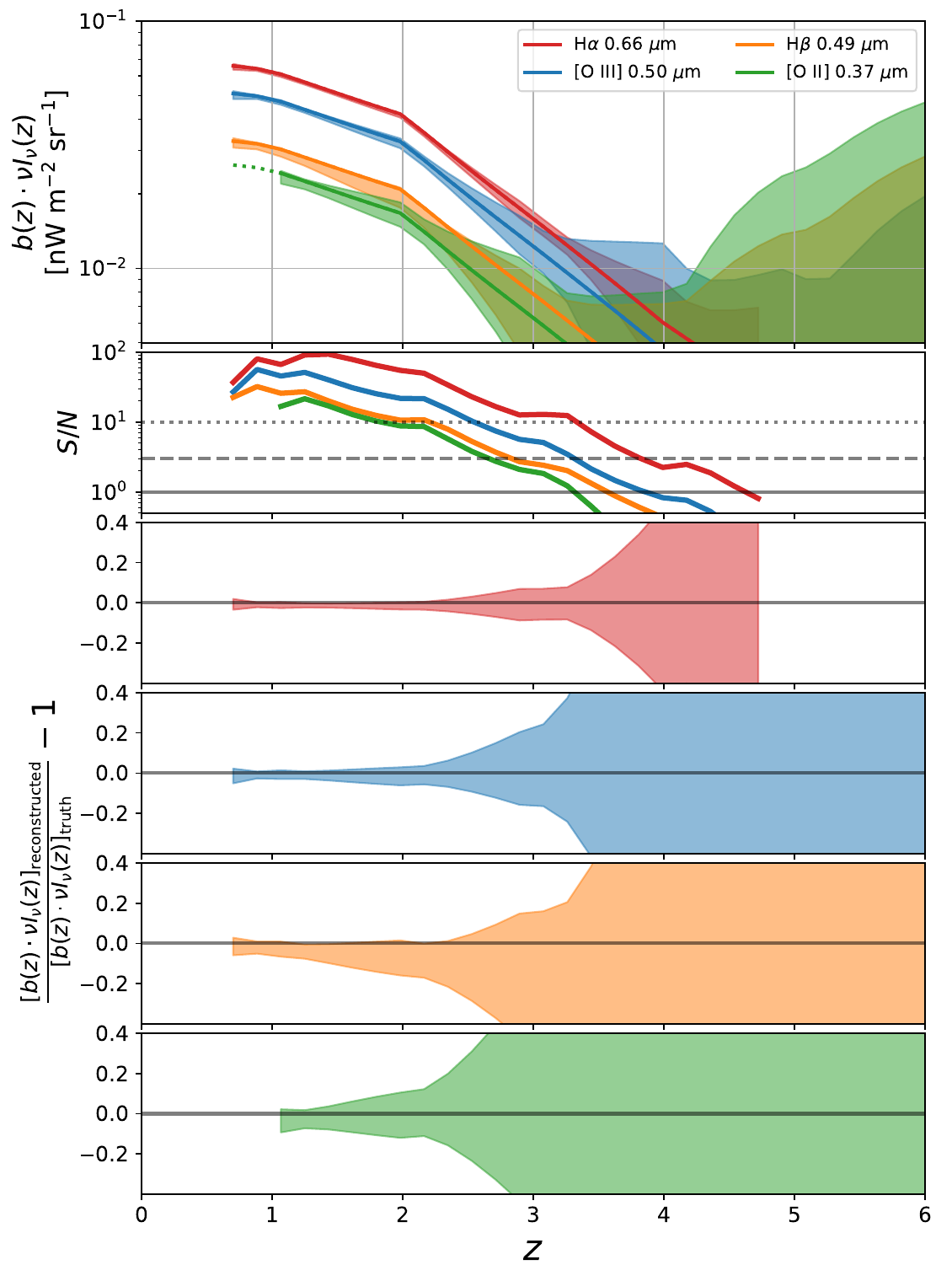}
\caption{\label{F:bnuInu_z_constraints} Top: our constraints on the bias-weighted intensity for each line from the fiducial case. The solid/dashed lines denote the input model within the redshift ranges accessible/inaccessible by the survey spectral coverage. We consider the line signal only from $0.7<z<6$ (see Section~\ref{S:survey_setup}). The colored shaded regions mark the 1$\sigma$ constraints from our inference with the case that contains sample variance in the input mock data (the orange case in Figure~\ref{F:corner}). Second panel: the S/N on the bias-weighted $b(z) \cdot \nu I_\nu(z)$ for each line (colored lines). The solid, dashed, and dotted black lines mark the 1$\sigma$, 3$\sigma$, and 10$\sigma$ sensitivity levels, respectively. Bottom four panels: 1$\sigma$ constraints for each line relative to the truth input (H$\alpha$, $[$\ion{O}{3}$]$, H$\beta$, and $[$\ion{O}{2}$]$, from top to bottom, respectively).}
\end{center}
\end{figure}

Next, we propagate our parameter constraints into the bias-weighted intensity for each line. We use Equation~\ref{E:cov_Mz} to first derive the constraints on $M_i(z)$, and we convert the $M_i(z)$ to intensity with Equation~\ref{E:nuInu_line}. The results are shown in Figure~\ref{F:bnuInu_z_constraints}. With our fiducial setup similar to the SPHEREx deep-field sensitivity, our algorithm can extract the bias-weighted intensity for all four lines at the $\gtrsim 10\sigma$ level at $z<2$. For the brightest line, H$\alpha$, the $10\sigma$ sensitivity can be achieved out to $z\sim3$, with the S/N (S/N) peaking at $\sim 100\sigma$ around $z=1.5$.
While in reality, the sensitivity depends on both the underlying signal model and the systematic uncertainties not accounted for in our forecast, our results suggest a promising prospect for simultaneously detecting LIM signals from multiple lines with SPHEREx, by leveraging the information encoded in the correlation between lines in the spectral--angular space.

\begin{figure}[ht!]
\begin{center}
\includegraphics[width=\linewidth]{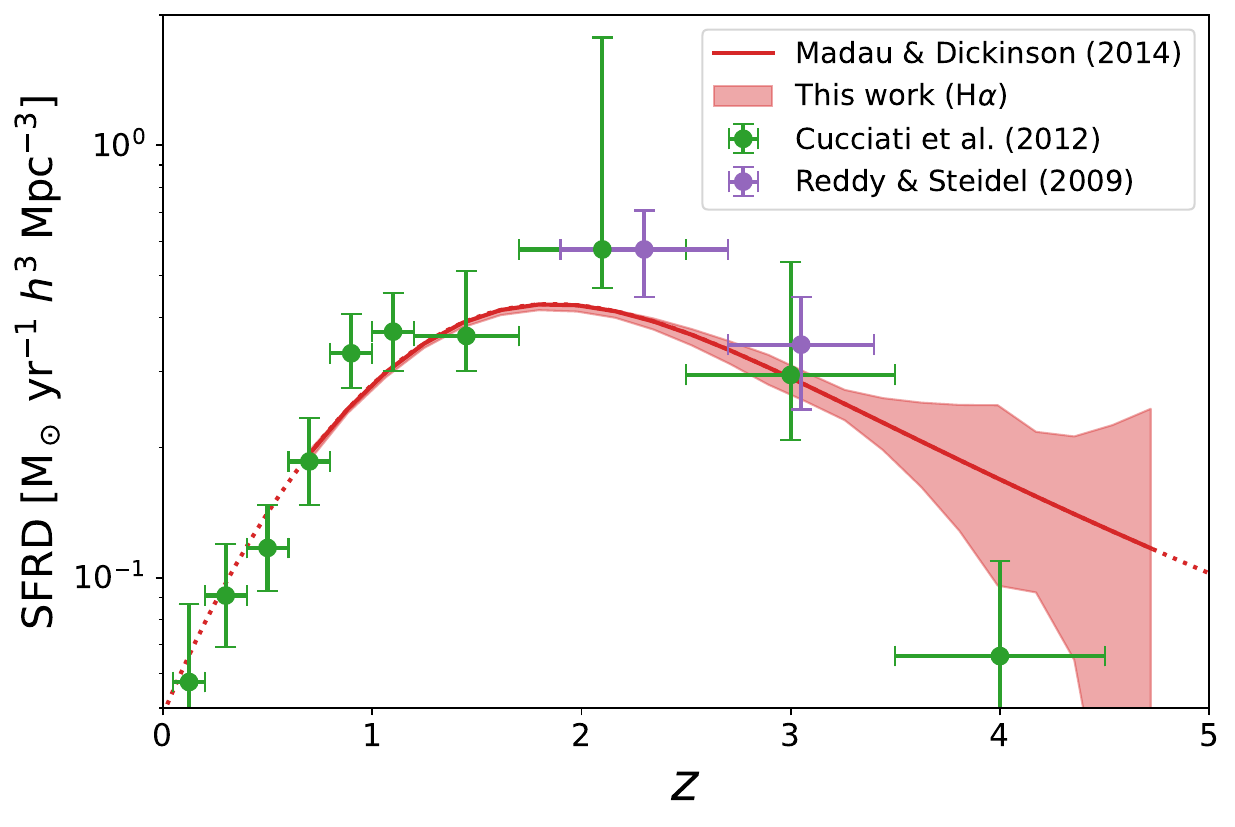}
\caption{\label{F:SFRD_constraints} The $1\sigma$ SFRD constraints from our fiducial case. We assume that the SFRD is proportional to the bias-weighted intensity of the H$\alpha$ line, and thus the sensitivity on the SFRD is propagated from our H$\alpha$ constraints, as shown in Figure~\ref{F:bnuInu_z_constraints} (red shaded region). The red curve denotes our underlying model for the SFRD from \citet{2014ARA&A..52..415M}. For comparison, we also display observational constraints from \citet[][; green]{2012A&A...539A..31C} and \citet[][; purple]{2009ApJ...692..778R}.}
\end{center}
\end{figure}

We can also estimate the constraining power on the SFRD using our constraints on the bias-weighted intensity. Here, we propagate our H$\alpha$ constraint to the SFRD, assuming that the bias-weighted intensity of H$\alpha$ is proportional to the SFRD and ignoring uncertainties in this conversion factor. In other words, we assume the same S/N for the bias-weighted intensity of H$\alpha$ and the SFRD. The results are shown in Figure~\ref{F:SFRD_constraints}. We see that with the SPHEREx survey setup and applying our algorithm for the LIM analysis, we can derive a competitive constraining power on the SFRD at $z\lesssim 3$. While our algorithm simultaneously infers multiple lines that trace the star formation history, allowing for a potentially tighter constraint on the SFRD through the joint information from all lines, this analysis will require additional modeling of the correlation between lines, which is beyond the scope of this work.

\section{Discussion}\label{S:discussion}

\subsection{Dependence on the Noise Level}

\begin{figure}[ht!]
\begin{center}
\includegraphics[width=\linewidth]{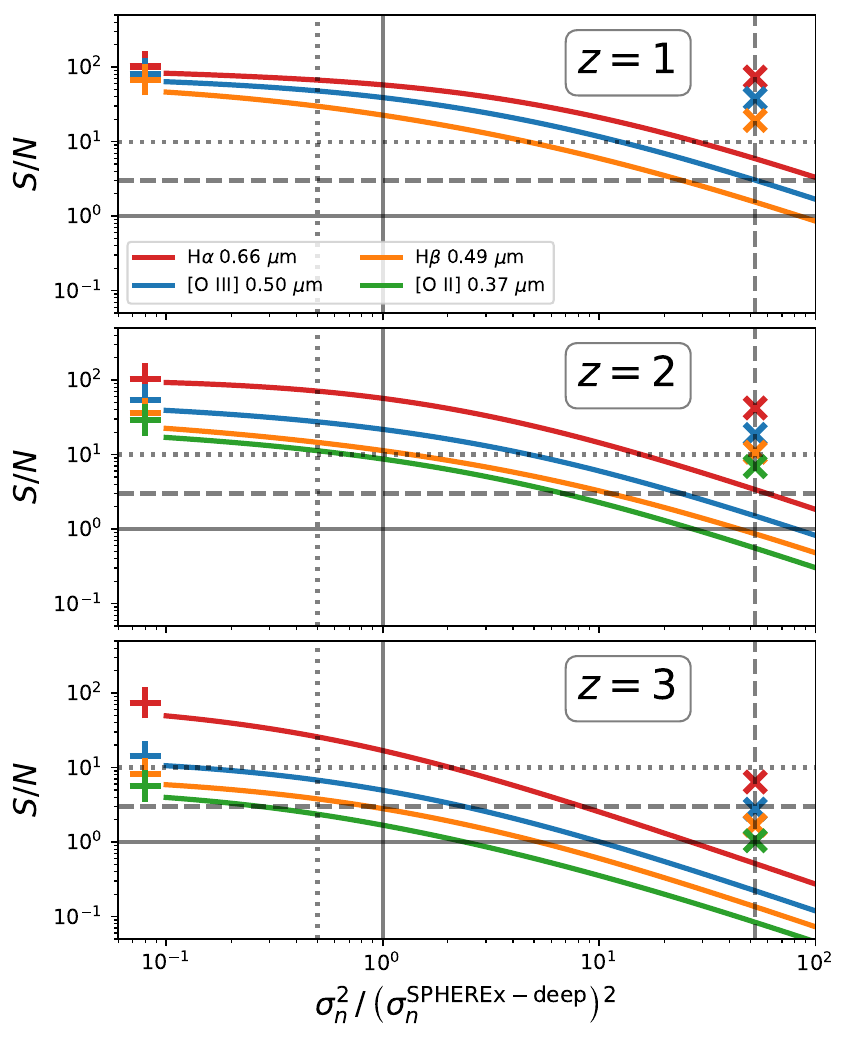}
\caption{\label{F:noise_scaling} The S/N on the bias-weighted intensity at $z=1$ (top), $2$ (middle), and $3$ (bottom) for each line is shown as a function of the noise level relative to the fiducial SPHEREx deep-field noise, $\sigma_n^{\rm SPHEREx-deep}$. The absence of $[$\ion{O}{2}$]$ at $z=1$ is because it falls outside the range that can be probed by our survey. The plus markers denote the noiseless case ($\sigma_n=0$). The solid, dashed, and dotted gray horizontal lines mark the 1$\sigma$, 3$\sigma$, and 10$\sigma$ sensitivity levels, respectively. The solid and dotted gray vertical lines mark $\sigma_n^2/(\sigma_n^{\rm SPHEREx-deep})^2=$1 and 0.5, respectively. The gray dashed vertical line marks the SPHEREx all-sky survey sensitivity. The colored crosses scale the all-sky S/N by the square root of the sky coverage ratio to represent the constraining power in the all-sky survey.}
\end{center}
\end{figure}

Our fiducial case, presented in Section~\ref{S:results}, considers the SPHEREx deep-field sensitivity. Here, we explore how the model constraints depend on the noise level. We use the same model for the line signal as in the fiducial case, apply different scaling to the fiducial noise variance in the SPHEREx deep field, $(\sigma_n^{\mathrm{SPHEREx-deep}})^2$, across all frequency channels, and use the Fisher matrix to derive parameter constraints as a function of the noise level.

The results are displayed in Figure~\ref{F:noise_scaling}. We observe that the sensitivity increases as the noise level decreases and approaches the noiseless limit (plus symbols) for bright lines at lower redshifts, where the line power is significantly stronger than the noise.

Figure~\ref{F:noise_scaling} also provides insights into the sensitivity of detecting the line signal if SPHEREx extends beyond its nominal 2 yr survey. For instance, if SPHEREx extends its mission lifetime to 4 yr, the noise variance will integrate down by a factor of 2, as indicated by the vertical dotted line in Figure~\ref{F:noise_scaling}. In this case, we find about a factor of 2 sensitivity improvement for detecting the line signals at $z=3$, whereas at $z=1$, the improvement is less evident, especially for brighter lines such as H$\alpha$, as the signals are already not in the noise-dominated regime, even in the case of the fiducial sensitivity of the nominal 2 yr mission.

Similarly, we can also estimate the sensitivity if we apply our algorithm to the SPHEREx all-sky survey instead of in deep fields. The SPHEREx all-sky noise variance is about 50 times higher than the deep field (Figure~\ref{F:spherex_R_noise}), which is denoted by the dashed vertical lines in Figure~\ref{F:noise_scaling}. While the curves in Figure~\ref{F:noise_scaling} indicate low sensitivity at this noise level, we note that we will have access to a much larger sky coverage in the all-sky survey. Fisher information on the parameters is proportional to the number of available modes and thus the sky coverage. Assuming SPHEREx all-sky coverage of $f_{\rm sky}^{\rm SPHEREx-all} \approx 75\%$, we get a factor of $\sim150$ more sky coverage than the deep field ($f_{\rm sky}=0.48\%$), and thus a $\sim12$ times boost of the S/N, indicated as the colored crosses in Figure~\ref{F:noise_scaling}. We find that at $z=1$, the all-sky and deep-field sensitivity is similar, while for higher redshift, at $z=3$, the deep field will perform better than the all-sky survey, as the higher-redshift line signals are fainter and thus more susceptible to the noise fluctuations.

We emphasize that we cannot achieve infinite sensitivity to the line signal, even in the absence of instrument noise (the plus symbols in Figure~\ref{F:noise_scaling}). This fundamental sensitivity limit is due to the nature of the line confusion in the data. In the spectral--angular space, emission from different lines is mixed together, and hence acts as ``line noise'' for each other.

\subsection{Presence of Correlated Noise}\label{S:corr_noise}
In this work, we assume that instrument noise is uncorrelated across channels, making the noise power spectra a diagonal matrix. In reality, the data usually contain correlated noise from the intrinsic instrumental noise, foreground residuals, and the continuum-filtering process.

\begin{figure}[ht!]
\begin{center}
\includegraphics[width=\linewidth]{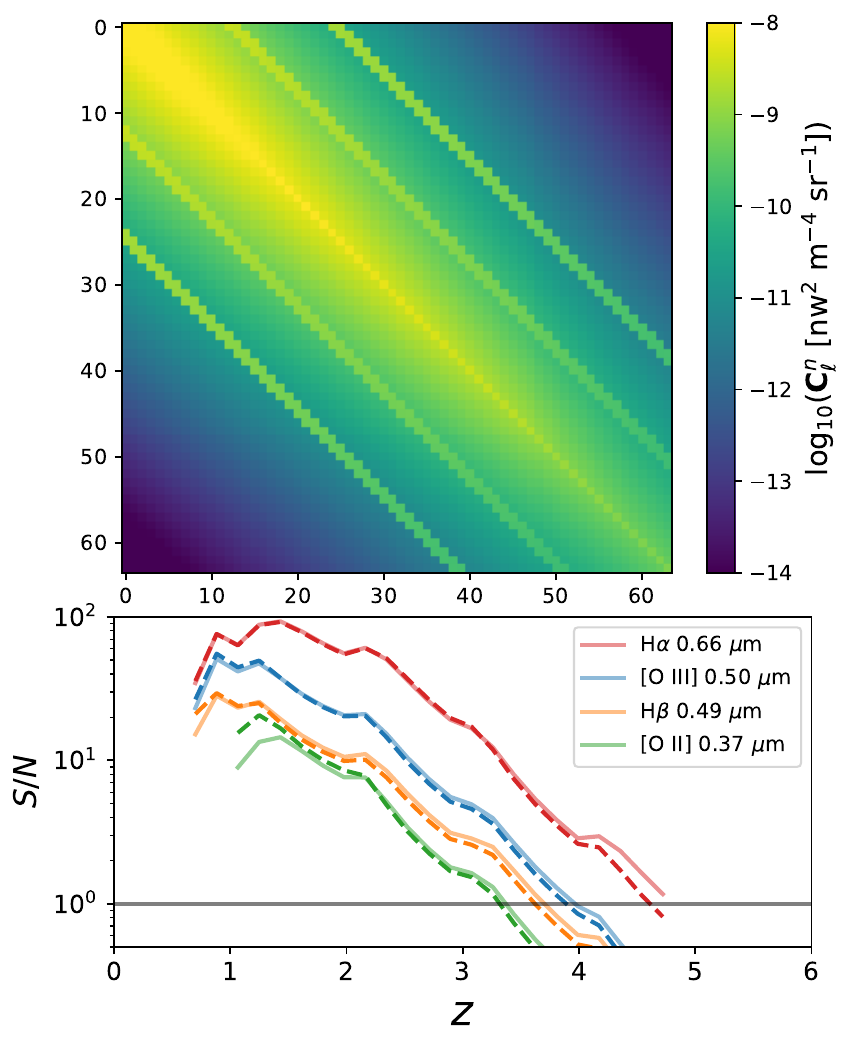}
\caption{\label{F:SNR_corr_noise} Top: an example of a correlated noise matrix. Bottom: comparison of the S/N on $M_i(z)$ for each line with uncorrelated (solid) and correlated (dashed) noise. We assume the same signal model and survey setup as the fiducial case, using the noise matrix shown in the top panel for the correlated noise and a diagonal matrix with the same eigenvalues for the uncorrelated noise.}
\end{center}
\end{figure}

To assess how correlated noise might affect the reconstruction results, we create a test case with correlated noise, as shown in the top panel of Figure~\ref{F:SNR_corr_noise}. We assume continuous correlated noise that decays with channel separation to emulate typical foreground residuals. Additionally, we add a few off-diagonal streaks that manifest similar features to line correlation signals. These kinds of noise features can be caused by, for example, detector crosstalk. We set the eigenvalues of this correlated noise matrix to be similar to our fiducial case.

We then quantify the S/N on $M_i(z)$ for each line with a Fisher forecast (Figure~\ref{F:SNR_corr_noise}, bottom panel). We compare the case of using the correlated noise shown in the top panel with the case of uncorrelated noise with the same eigenvalues. We find the parameter constraints to be almost identical in both cases\footnote{We note that marginalized constraints on individual parameters depend nontrivially on the noise covariance, and therefore the S/N for certain parameters in the presence of correlated noise could be higher than in the noncorrelated case.}. This indicates that our algorithm is not strongly affected by the presence of correlated noise. This is because the correlated line signal has a certain (and deterministic) pattern in the cross-power-spectrum space, so only correlated noise that manifests the exact pattern of the spectral correlation of the signal will induce significant degeneracy in our inference.

\subsection{Information from Small Scales}\label{S:ellmax2000}
\begin{figure}[ht!]
\begin{center}
\includegraphics[width=\linewidth]{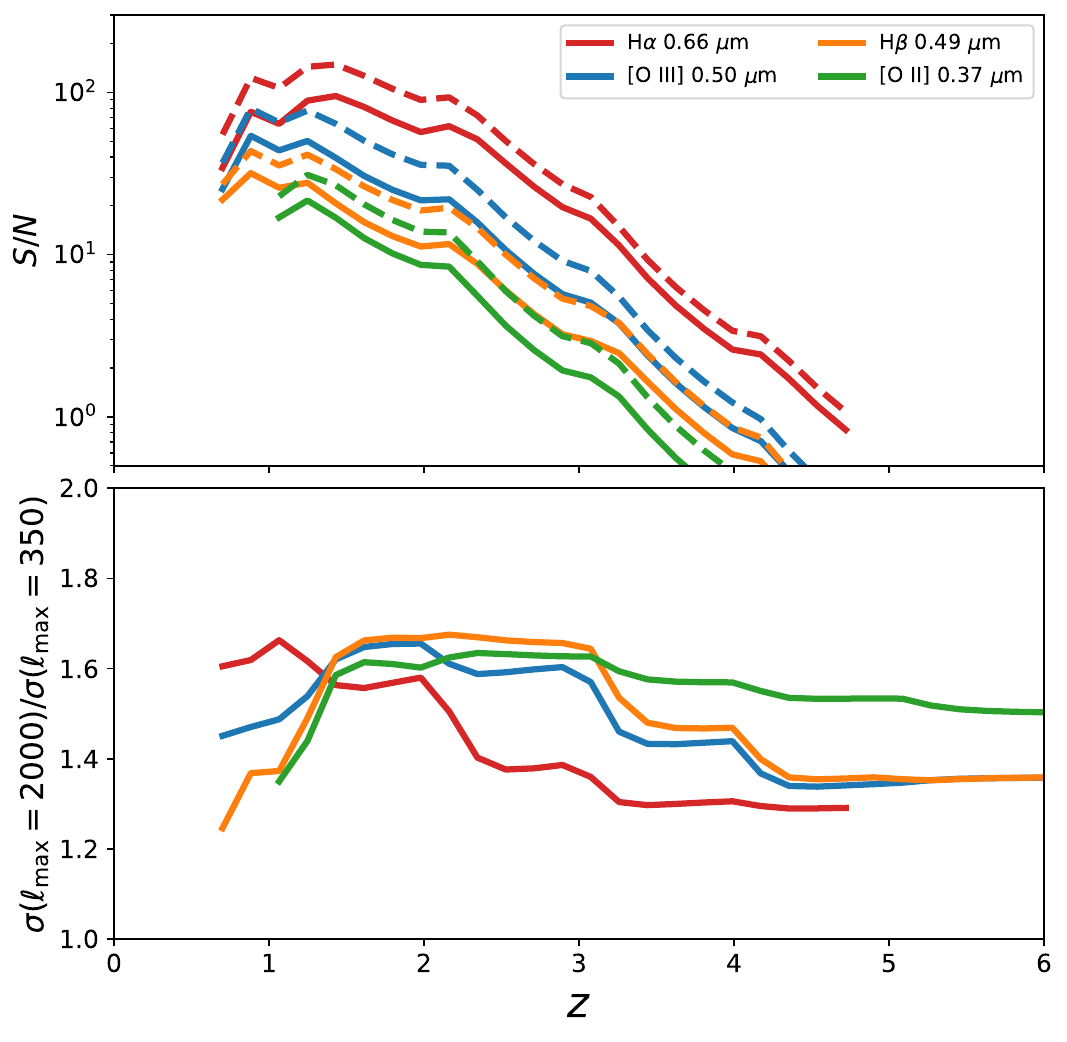}
\caption{\label{F:SNR_ellmax2000} Top: the S/N of line reconstruction in the fiducial setup (solid) and the case with $\ell_{\rm max}$ extending from $350$ to $2000$ (dashed). Bottom: the ratio of the inferred line sensitivity of the extended $\ell_{\rm max}$ case and the fiducial case.}
\end{center}
\end{figure}

Our fiducial setup limits the analysis to large scales ($\ell < 350$) to model the signal from only the linear regime. This discards a huge amount of smaller-scale modes accessible by SPHEREx. To assess the information content from these higher-$\ell$ modes, we calculate the S/N on line reconstruction with a setup that extends the maximum multipole mode to $\ell_{\mathrm{max}}=2000$, while keeping other assumptions the same as the fiducial model. The Poisson noise becomes non-negligible on smaller scales; therefore, in this calculation, we include a model of Poisson noise (Appendix~\ref{A:SN}) for both cases. The results are shown in Figure~\ref{F:SNR_ellmax2000}. While there are $\sim 30$ times more modes between $350 < \ell < 2000$ compared to our fiducial setup of $50<\ell<350$, the higher-multipole modes correspond to higher-$k$ modes in the matter power spectrum $P(k)$ with a lower power and thus are more susceptible to the (white) instrument noise, and thus they do not contain as much information as the lower-$\ell$ modes. Therefore, there is only a factor of $\sim1.6$ gain in the line sensitivity by the inclusion of the higher-$\ell$ modes. We also note that in reality, to extract information from small scales, one must include the effect of nonlinear clustering in the model, which we have ignored in this calculation.

\begin{figure}[ht!]
\begin{center}
\includegraphics[width=\linewidth]{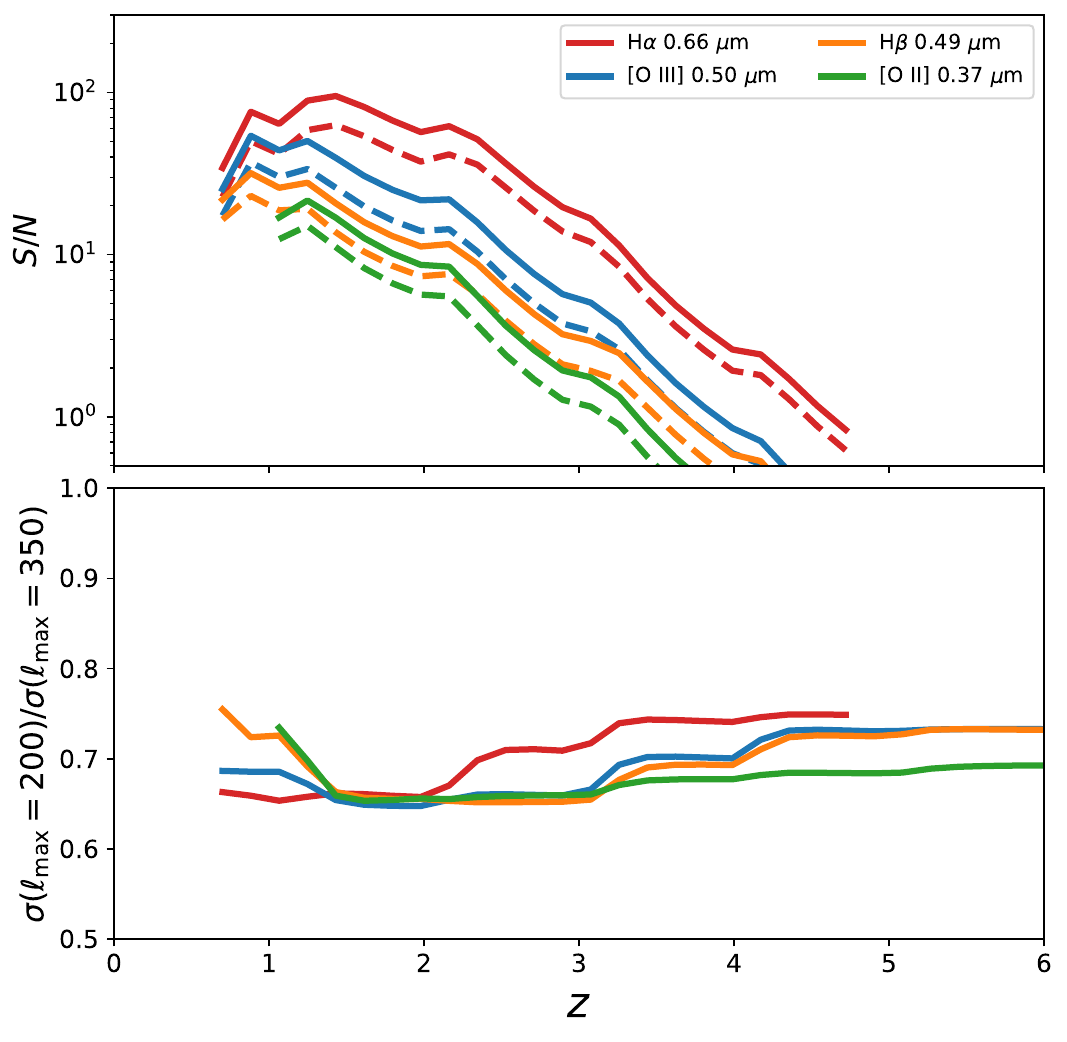}
\caption{\label{F:SNR_ellmax200} Top: the S/N of line reconstruction in the fiducial setup (solid) and the case with $\ell_{\rm max}$ reducing from $350$ to $200$ (dashed). Bottom: the ratio of the inferred line sensitivity of the extended $\ell_{\rm max}$ case and the fiducial case.}
\end{center}
\end{figure}

Similarly, we can estimate the information loss if higher-multipole modes have to be discarded due to, for example, higher nonlinear bias of the line emission field compared to our current model. Figure~\ref{F:SNR_ellmax200} compares the S/N when reducing $\ell_{\rm max}$ from $350$ to $200$. While this corresponds to discarding $70\%$ of multipole modes, the sensitivity on $M_i(z)$ is only reduced by $\sim30\%$, for the same reason as above: the large-scale modes are less noisy than the small-scale modes.

\subsection{Capability of Interloper Separation}\label{S:Pell}
\begin{figure*}[ht!]
\begin{center}
\includegraphics[width=\linewidth]{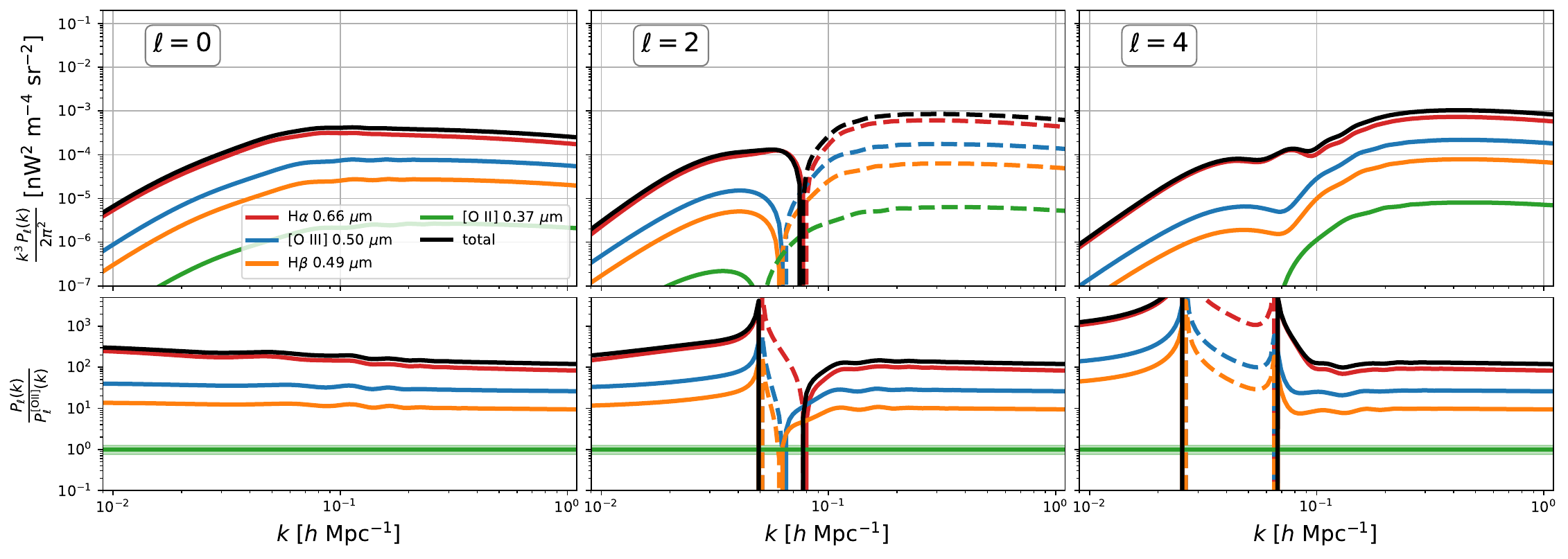}
\caption{\label{F:Pell_constraints_OII_z2.5} Top: the 3D power spectrum multipoles $P_\ell(k)$ of the lines projected to the $[$\ion{O}{3}$]$ frame at $z=2.5$. We show the first three multipole modes at $\ell=0$ (left), $\ell=2$ (middle), and $\ell=4$ (right).  Bottom: the ratio of $P_\ell(k)$ from each line to the target line $[$\ion{O}{3}$]$. The shaded region on the $[$\ion{O}{3}$]$ line (green) denotes the 1$\sigma$ constraint of the bias-weighted intensity ($b_i(z)\nu I_\nu(z)$) from our fiducial case (Section~\ref{S:results}), which gives a S/N of $3.9$ at $z=2.5$.}
\end{center}
\end{figure*}

Our method jointly constrains signals from multiple spectral lines in LIM datasets. Here, we highlight the capability of our method in extracting faint line signals from multiple interlopers with orders of magnitude stronger power. As the most commonly used summary statistic in LIM is the 3D power spectrum $P(k)$, we present our model constraints in this representation. Specifically, we show the first few multipole moments of $P(k)$, which capture the anisotropy of the line power spectrum. This anisotropy is due to the incorrect projection of interloper lines to the target line redshift. In LIM analysis, the $i$th interloper signal will be projected from the redshift $z_i$ to the target line redshift $z_t$, where $z_i = \lambda_t(1 + z_t)/\lambda_i - 1$, making the interloper and the target line fall in the same observed frequency. The different projection in the transverse and the LOS directions makes the interloper power spectrum anisotropic \citep{2016ApJ...825..143L, 2016ApJ...832..165C, 2020ApJ...894..152G}, which introduces nonzero $\ell>0$ modes when expanding the 3D power spectrum with Legendre polynomials \citep{2019PhRvD.100l3522B, 2020ApJ...894..152G, 2021PhRvD.103f3523B}.

The $\ell$th multipole of the total LIM power spectrum is the sum of the contribution from all lines:
\begin{equation}
P_\ell(k,z_t) = \sum_{i=1}^{N_{\rm line}} P_\ell^i(k,z_t).
\end{equation}
The $P_\ell$ from the $i$th line is given by
\begin{equation}\label{E:Pell_i}
\begin{split}
P_\ell^i(k,z_t) &= \frac{1}{\left(q^i_\perp\right)^2 q^i_\parallel} \,\frac{2\ell+1}{2}\\
&\cdot\int_{-1}^{1} d\mu \,P^i(k_i,\mu_i,z_i)\, \mathcal{L}_\ell(\mu),
\end{split}
\end{equation}
where $\mathcal{L}_\ell$ is the Legendre polynomial, $q^i_\perp$ and $q^i_\parallel$ are the projection factors in the transverse and the LOS directions, respectively, and $P^i(k_i,\mu_i,z_i)$ is the intrinsic power spectrum of the $i$th line at redshift $z_i$ and the corresponding Fourier mode $k_i$ and cosine angle $\mu_i$. For the intrinsic power spectrum, $P^i(k_i,\mu_i,z_i)$, we also incorporate the RSD effect \citep{1987MNRAS.227....1K} and the window functions due to the finite resolution in the LOS and transverse directions. Both effects introduce additional sources of anisotropy to the observed power spectrum. The full expression of these quantities is presented in Appendix~\ref{A:Pell}.

Figure~\ref{F:Pell_constraints_OII_z2.5} presents the power spectrum multipoles $P_\ell^i(k,z_t)$ for the three lowest-$\ell$ modes. \footnote{Since the line power spectrum $P^i(k_i,\mu_i,z_i)$ is an even function of the LOS cosine angle $\mu$, the odd multipole modes vanish.} Here, we choose $[$\ion{O}{2}$]$ as the target line and present the results at the target line redshift $z_t=2.5$, corresponding to the observed wavelength at $1.3$ $\mu$m, and the interlopers are from redshifts $z_i=$ $0.99$ (H$\alpha$), $1.61$ ($[$\ion{O}{3}$]$), $1.68$ (H$\beta$).

From Figure~\ref{F:Pell_constraints_OII_z2.5}, we see that the $[$\ion{O}{2}$]$ power spectrum is overwhelmed by interloper power by more than 2 orders of magnitude in all three multipole modes. Nevertheless, our method makes use of the information on the frequency correlations between lines and can successfully extract the bias-weighted intensity of $[$\ion{O}{2}$]$ at $z_t=2.5$ with an S/N of $3.9$.

\subsection{Robustness Against Model Misspecification}\label{S:model_misspecification}

\begin{figure}[ht!]
\begin{center}
\includegraphics[width=\linewidth]{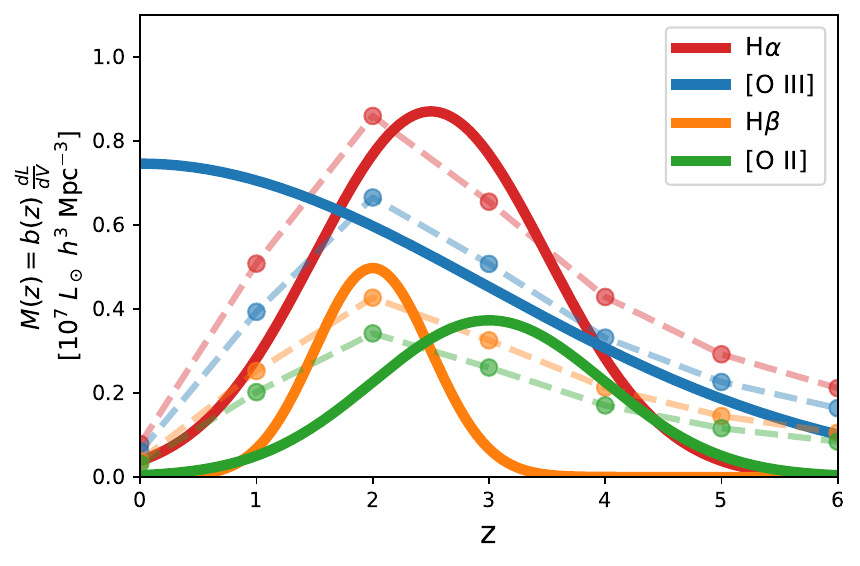}
\caption{\label{F:Mz_model_misspec}The biased-weighted luminosity density for the four spectral lines in our test case for model misspecification (solid lines). The dashed lines are our fiducial case for comparison. The fiducial case assumes a piecewise linear function with anchoring redshifts at $z=$0,1,...,6 (points on dashed lines).}
\end{center}
\end{figure}

All calculations up to this point have utilized the input line signal from our fiducial model (Section~\ref{S:fiducial_params}), constructed from our piecewise linear parameterization. While the piecewise linear function offers great flexibility to approximate any continuous function, with a limited number of anchoring points, it cannot perfectly capture realistic bias-weighted luminosity density functions, which are expected to be smooth curves. To validate that our algorithm can faithfully reconstruct signals with a different underlying input, we perform the following test. We generate the input mock data $\{\mathbf{C}_\ell^d\}$ using Gaussian functions for the bias-weighted luminosity densities $M_i(z)$ for each line, as illustrated in Figure~\ref{F:Mz_model_misspec}. In addition to testing the robustness of the reconstruction under our piecewise linear approximation, we also intentionally assign very different shapes of $M_i(z)$ for each line compared to the fiducial case, to assess whether our algorithm can still reconstruct the input accurately.

\begin{figure}[ht!]
\begin{center}
\includegraphics[width=\linewidth]{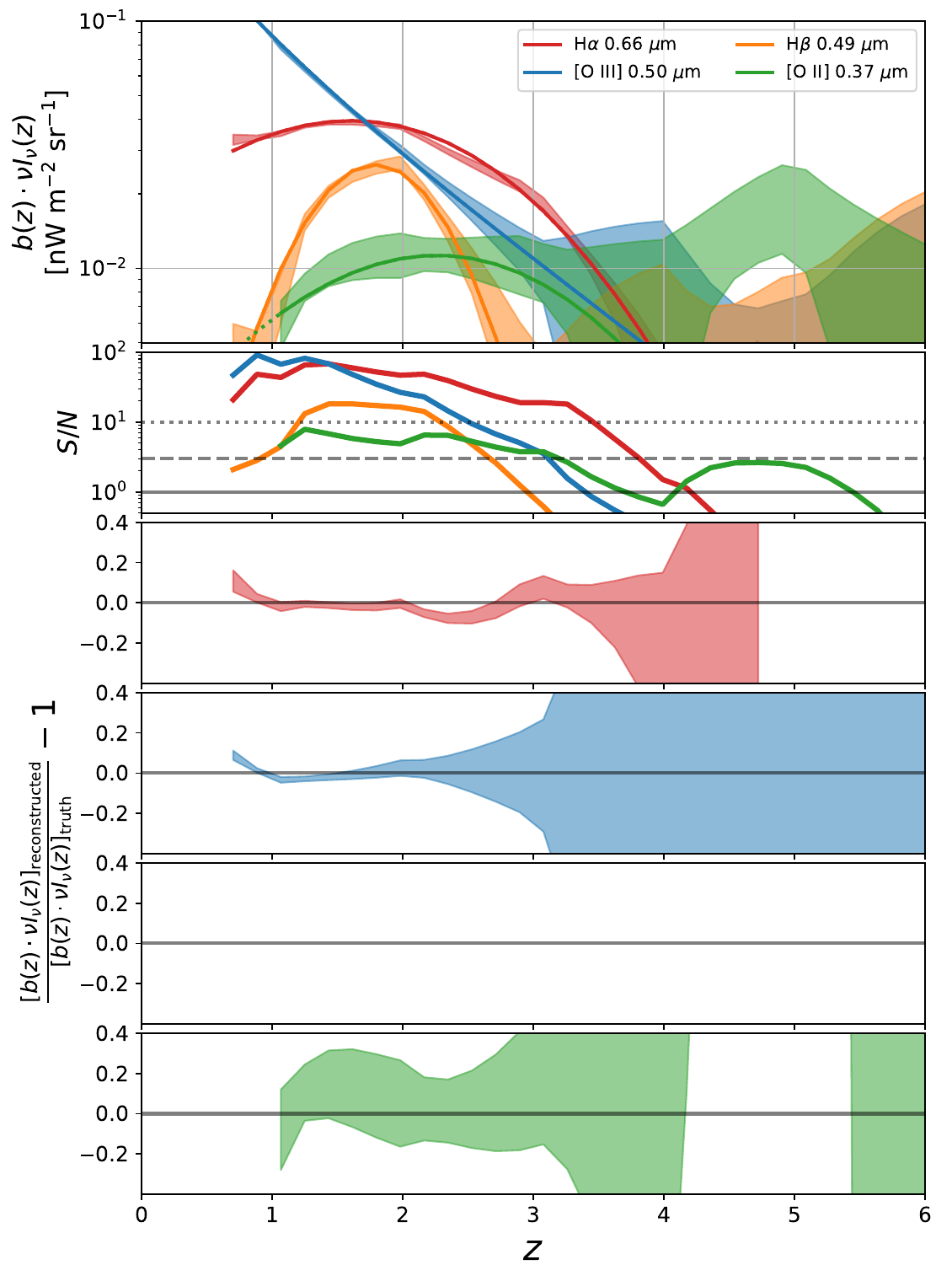}
\caption{\label{F:bnuInu_z_constraints_model_misspec_dz1} Top: our constraints on the bias-weighted intensity for each line, where the line signal in the input mock data is shown in Figure~\ref{F:Mz_model_misspec}, and we reconstruct the signal with our piecewise linear parameterization. The solid/dashed lines denote the input model within the redshift ranges accessible/inaccessible by the survey spectral coverage. The colored shaded regions mark the 1$\sigma$ constraints from our inference. The input mock data contain sample variance fluctuations. Second panel: the S/N on $b(z) \cdot \nu I_\nu(z)$ for each line (colored lines). The solid, dashed, and dotted black lines mark the 1$\sigma$, 3$\sigma$, and 10$\sigma$ sensitivity levels, respectively. Bottom four panels: 1$\sigma$ constraints for each line relative to the truth input (H$\alpha$, $[$\ion{O}{3}$]$, H$\beta$, and $[$\ion{O}{2}$]$, from top to bottom, respectively).}
\end{center}
\end{figure}

\begin{figure}[ht!]
\begin{center}
\includegraphics[width=\linewidth]{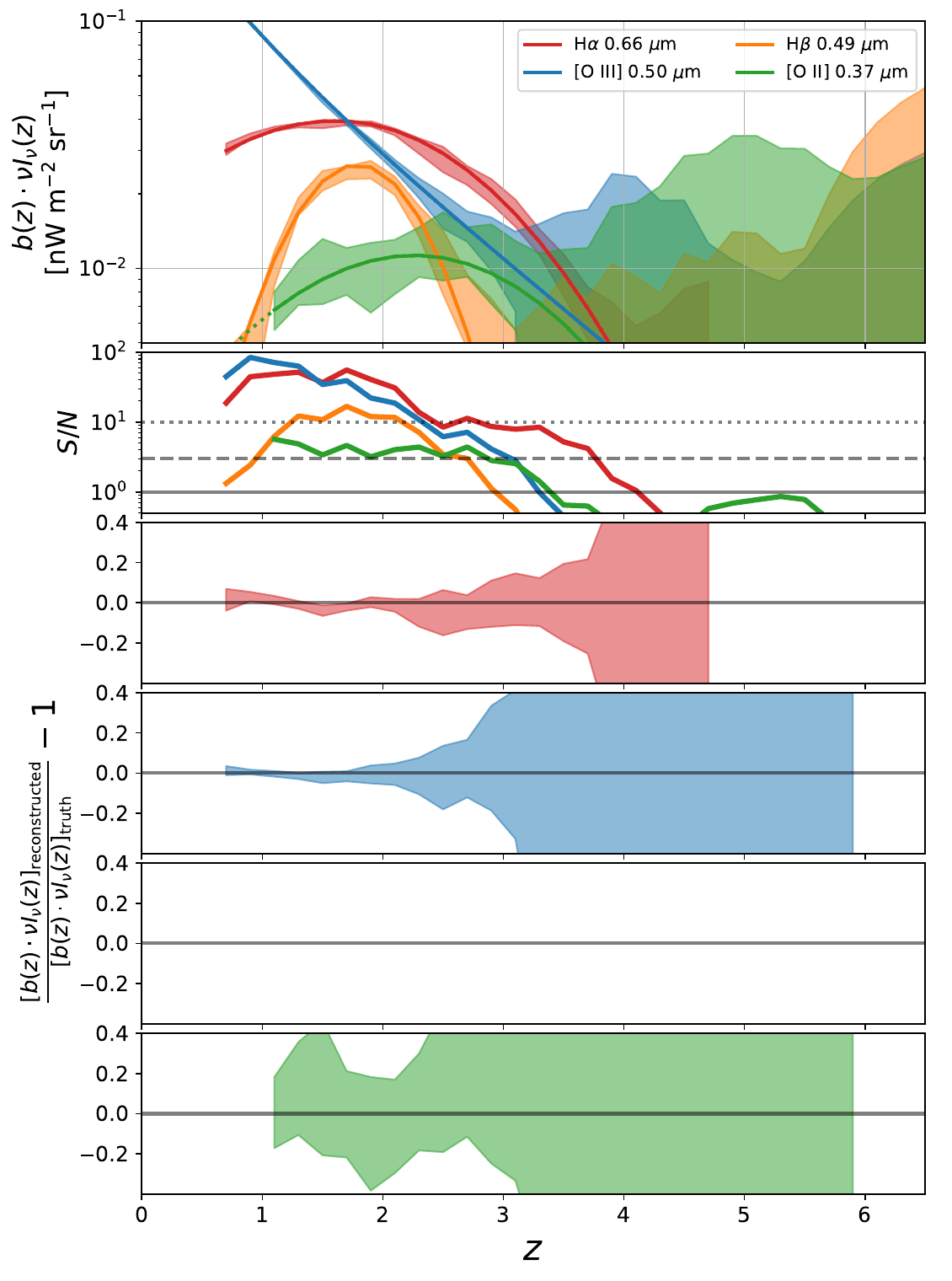}
\caption{\label{F:bnuInu_z_constraints_model_misspec_dz05} Reconstruction results of the same input signal as in Figure~\ref{F:bnuInu_z_constraints_model_misspec_dz1} with finer redshift anchoring points of $\Delta z=0.5$ instead of the fiducial case of $\Delta z=1$ (Figure~\ref{F:bnuInu_z_constraints_model_misspec_dz1}).}
\end{center}
\end{figure}

Then, we apply our algorithm to infer the line signal from the data using our parameterization, i.e., approximating the input $M_i(z)$ with the piecewise linear function. 
The results are shown in Figure~\ref{F:bnuInu_z_constraints_model_misspec_dz1}. We find that even if our model cannot perfectly describe the true signal, our inference can still unbiasedly reconstruct the signals within a $\sim 1\sigma$ range at $z<4$. Beyond this redshift, however, we obtain biased results for the faintest line, $[$\ion{O}{2}$]$, at $z\sim5$. We checked that this type of bias depends on the specific realization of the sample variance. To further investigate this feature, we ran the inference on the same data power spectra $\{\mathbf{C}_\ell^d\}$ with doubled redshift resolution, i.e., setting the redshift anchoring point spacing to $\Delta z=0.5$ instead of the fiducial case of $\Delta z=1$. The results are shown in Figure~\ref{F:bnuInu_z_constraints_model_misspec_dz05}. We find a higher best-fit likelihood value, and the inference on $M_i(z)$ is less biased than the fiducial case. This indicates that, in practice, it is essential to optimize the trade-off between the complexity and flexibility of our model in determining the number of redshift parameters. We leave more detailed analysis of this optimization to future works.

\subsection{Implementation with Continuum Foregrounds}\label{S:continuum}

Our model ignores continuum foregrounds that will be present in LIM data in practice. The continuum includes both Galactic and extragalactic emission. The continuum usually has a smooth spectrum, which we assume is already being filtered out in the map space before computing the cross-power spectra $C_{\ell,\nu\nu'}$. There are extensive studies on continuum foreground removal strategies for LIM---for example, principal component analysis \citep[PCA;][]{2008MNRAS.388..247D, 2010Natur.466..463C, 2013ApJ...763L..20M, 2013MNRAS.434L..46S, 2023A&A...676A..62V}, the asymmetric re-weighted penalized least squares \citep[arPLS;][]{2023A&A...676A..62V}, and semi-blind component separation techniques based on independent component analysis \citep[ICA;][]{2016ApJS..222....3Z}.

Part of the extragalactic continuum is emitted by the same galaxies that emit the line signal. Thus, a joint framework that simultaneously fits for lines and continuum can provide additional constraints on both galaxy physics and the underlying LSS. We defer this investigation to future studies.

\section{Advantages of Our Method}\label{S:unique_advantages}
\subsection{Multiline Inference across Redshifts}
Our method infers the bias-weighted line intensity as a function of redshift from multiple lines. This differs from many previously proposed LIM analysis techniques that operate in the 3D power spectrum ($P(k)$) space, approximating the line signal from different frequencies as originating from the same redshift (detailed in Section~\ref{S:3D_xPS}). The exceptions are some works that investigate the ``antisymmetric cross power spectrum'' between the CO and \ion{H}{1} lines during the EoR \citep{2020PhRvD.102d3519S,2021ApJ...909...51Z}. These studies show that the cross-correlation of the two lines at slightly offset LOS distances is not symmetric under exchange, yielding a non-zero antisymmetric cross-spectrum that contains additional information to constrain the EoR parameters. We emphasize that while we adopt the Limber approximation, which ignores any LOS correlation in this work, the full expression of the $C_{\ell,\nu\nu'}$ matrices will incorporate this information, as the redshift evolution model has been encoded in our formalism.

Additionally, some line deconfusion techniques treat interlopers as nuisance emission and seek strategies to mitigate interloper contamination in order to detect the target line. Here, our analysis is not restricted to extracting only one target line, allowing us to jointly reconstruct the signal from multiple spectral lines.

The intensity from multiple lines can provide a wealth of information. For example, many atomic or molecular lines serve as good tracers for the star formation history, and combining constraints from multiple lines can enhance the understanding of the global SFRD evolution. The ratio between the line signals also offers a valuable astrophysical census. For instance, given that the intrinsic ratio between H$\alpha$ and H$\beta$ luminosity is fixed by atomic physics ($L_{H\beta}/L_{H\alpha}=0.35$), deviations from this ratio probe the dust attenuation law. However, extracting these astrophysical constraints requires breaking the degeneracy between bias and intensity, either by using ancillary information or by jointly modeling the bias and intensity based on more detailed simulations.

\subsection{Straightforward Implementation}
Our model uses the auto-/cross-spectra between spectral channels, which is the covariance of the native form of the LIM data product. Many instrumental effects and foregrounds are naturally being described in this spectral--angular space (e.g., the filter transmission profile, noise correlations, atmospheric emission). Although, in this work, we disregard these components in our analysis, our technique provides a more compatible framework for incorporating these effects in reality.

In contrast, many other LIM analysis methods convert the data into the comoving space of the target line and compute the 3D power spectrum $P(k)$. This conversion relies on the assumed cosmological model, and furthermore, systematics may arise from 
errors in interpolation and projection \citep{2023arXiv231207289C}.

\subsection{Flexibility}
Our flexible framework can accommodate any form of the signal model. In this study, we choose a piecewise linear model to fit the bias-weighted luminosity density $M_i(z)$, providing good flexibility in approximating various functional forms. However, any parameterization for the signal can be implemented within our framework, as long as we can express the likelihood's dependence on the parameters. Additionally, incorporating any prior assumptions is straightforward, either through designing the parameterization of $M_i(z)$ or encoding them into the Bayesian prior. While we only apply a positivity prior on $M_i(z)$ in this work, one can introduce other prior information, such as a prior on the line ratio between certain pairs of lines.

Moreover, despite fixing the cosmological model in this analysis, one can also simultaneously fit for cosmological information and the line signal. This approach was demonstrated in C23, where we jointly fit for the matter power spectrum $P(k)$ and the spectral and redshift dependence of the emission.

\subsection{Generalizability} 
While we demonstrate signal reconstruction using cross-channel correlations within the same LIM dataset, our framework can be extended to include correlations with other datasets. For instance, we can cross-correlate with other LIM surveys and   photometric/spectroscopic galaxy catalogs to derive joint constraints from multiple probes.

\section{Comparison with Other LIM analysis Methods}\label{S:compare_previous_works}

In this section, we summarize the main differences of our technique and a few other LIM analysis methods in the literature.

\subsection{3D Cross-power Spectrum}\label{S:3D_xPS}
Several analyses have investigated the detectability of cross-correlation between different lines, either within the same LIM dataset or with different experiments that probe lines within the same cosmic volume \citep{2010JCAP...11..016V,2011JCAP...08..010V,2012ApJ...745...49G,2016ApJ...833..153S,2023arXiv231208471R}. Multiple line--line cross-spectra can also be used to reconstruct the auto-spectra of the target line \citep{2019ApJ...874..133B,2021JCAP...05..068S,2023arXiv230800749M}. These analyses consider cross-correlation on the 3D power spectrum $P(k)$, requiring the projection of LIM data into comoving space and assuming the projected line signal at a fixed redshift. In contrast, our method does not involve projection before computing the cross-power spectrum, allowing us to model the redshift evolution of the line signal instead of assuming the emissions are from a single redshift.

The 3D power spectrum provides a simpler basis for extracting information from the LOS modes, which are useful for breaking the bias and intensity degeneracy from the RSD effect. The LOS correlations of the underlying density fluctuations are ignored in this work, due to the assumption of a top-hat window function and the Limber approximation. Without these simplifications, our formalism could also extract the LOS correlation. However, modeling LOS modes in the angular correlation space requires a double-Bessel-function integration (Equation~\ref{E:Cl_line_bessel}), whereas in the 3D power spectrum space, the signal is simply a Fourier transform of the field. Therefore, our frequency angular correlation and the 3D power spectrum are suitable for different analysis purposes and will be complementary to each other in practice.

\subsubsection{Angular Power Spectrum Covariance}\label{S:Cl_cov}
\citet{2019ApJ...875...86F} also employ a Bayesian framework to analyze the auto-/cross-frequency power spectrum in LIM in the context of component separation for the cosmic near-infrared background. However, they approximate the likelihood on the $C_{\ell,\nu\nu'}$'s as a Gaussian distribution, which is only valid in the limit of high S/N and with a large number of modes \citep{2008PhRvD..77j3013H}. In contrast, in our analysis, our data vector is the native LIM data product---the voxel intensity map (in the spherical harmonic space), with a covariance matrix given by $C_{\ell,\nu\nu'}$, which can fully capture the information from the data at the two-point level. 

Furthermore, the analysis in \citet{2019ApJ...875...86F} fits the observed power spectra with a set of amplitudes associated with predefined theoretical templates, making it more susceptible to model misspecification. In contrast, our framework employs flexible parameterization, allowing it to accommodate signals that might not be well described by existing models.

\subsubsection{Power Spectrum Anisotropy}
The anisotropy of the 3D power spectrum $P(k)$ of interloper lines, upon projection to the target line redshift (Section~\ref{S:Pell}), has been proposed as a strategy for line separation in LIM studies \citep{2016ApJ...825..143L,2016ApJ...832..165C,2020ApJ...894..152G}. However, this technique encounters difficulties with interloper lines that have rest-frame frequencies very close to that of the target line (such as H$\beta$ and $[$\ion{O}{3}$]$), due to the minimal effect of projection. In contrast, our method, which utilizes the complete information available in the spectral--angular space, is capable of successfully extracting the signals from all lines, even in cases where there is a closely located interloper.



\subsubsection{Pixel-space Spectral Template Fitting}
\citet{2020ApJ...901..142C} introduce a technique to extract the intensity map from individual lines in LIM, also using cross-frequency information. By fitting the spectrum in each LIM pixel to a large dictionary of spectra that encode which frequency channel contains line emission for a given redshift, they can successfully infer the redshift of the line emitters that are brighter than the noise level and thus reconstruct the line intensity map from those sources.

Our method, while not capable of reconstructing the individual line signal in the map space, makes use of the emission from all sources, rather than only detectable sources. Therefore, it is not restricted to the relatively low-noise and low-confusion regime as in \citet{2020ApJ...901..142C}. Furthermore, our angular power spectrum utilizes both the spectral correlations from multiple lines and the spatial correlation of the underlying cosmological field, whereas the pixel-by-pixel fitting in \citet{2020ApJ...901..142C} does not involve spatial clustering information.

\subsubsection{Machine Learning}
Machine learning (ML) has shown promise in decomposing emissions from different lines in LIM \citep{2020MNRAS.496L..54M,2021ApJ...923L...7M}, leveraging information beyond the two-point statistics that is not captured in the power spectrum. However, the effectiveness of ML approaches hinges on the availability of a comprehensive training set that covers the full range of potential signals and systematic variations. This requirement is challenging for many current LIM experiments, given our limited understanding of the underlying signal models, foreground, and instrument systematics.

\section{Future Work}\label{S:future_works}
Here, we outline directions toward better realism and broader applications of our technique for future studies.

Our previous work of C23 (continuum) establishes the inference algorithm for constraining continuum emission with broadband photometry from the frequency--frequency cross-correlation, while this study applies a similar framework for the line emission. A joint inference with both continuum and line emission will exploit the information from the full SEDs of galaxies. 

Furthermore, while we only perform inference on cross-frequency correlations ($C_{\ell,\nu\nu'}$), our framework can also be extended to incorporate cross-correlation with other tracers, such as photometric or spectroscopic galaxies \citep{2022ApJ...925..136C}.

Finally, Our analysis framework can also serve as a tool to search for unknown spectral features that trace the LSS (Cheng et al. 2024 in prep.). This is of great interest in searching for dark matter candidates decaying into photons that may manifest as unexpected lines in the LIM data \citep{2018PhRvD..98f3524C,2021PhRvD.103f3523B}.

\section{Conclusion}\label{S:conclusion}
In this work, we introduce a novel technique for analyzing LIM data with multiple line signals. While the presence of multiple line signals in LIM poses a challenging analysis issue, known as interloper contamination, our method leverages the correlated information from multiple lines to perform joint inference on all lines simultaneously. This is enabled by the correlated signal from lines originating from the same redshift, manifesting as unique off-diagonal signals in the covariance of the LIM data ($C_{\ell,\nu\nu'}$'s).

We employ Bayesian analysis to infer the bias-weighted intensity of each line from the data covariance $C_{\ell,\nu\nu'}$'s. Without relying on any external dataset, and only making use of assumptions of the signal homogeneity and isotropy, as well as the positivity condition on the bias-weighted intensity of all lines, our method enables the full exploitation of the information in the data on large scales, where the line emission field is Gaussian and can be fully characterized by two-point statistics.

We apply our method to mock LIM power spectra generated from a survey setup similar to the SPHEREx deep-field observation, considering four lines within the SPHEREx spectral coverage: H$\alpha$, $[$\ion{O}{3}$]$, H$\beta$, and $[$\ion{O}{2}$]$. We demonstrate that our algorithm can constrain the bias-weighted intensity of all four lines at the $\gtrsim 10\sigma$ level at $z<2$. For the brightest line, H$\alpha$, the $10\sigma$ sensitivity can be achieved out to $z\sim3$, with the S/N peaks at $\sim 100\sigma$ around $z=1.5$. We also show that our method is robust against model misspecification.

This work lays the foundation for broader applications in analyzing LIM data with various spectral features, which is timely, as many LIM experiments are expected to come online in the near future.

\section*{acknowledgements}
We would like to thank the anonymous referee for valuable comments that improved the manuscript. We are grateful to Asantha Cooray, Richard Feder, Adam Lidz, Jordan Mirocha, and Mike Zemcov for constructive discussions and comments on a draft manuscript, as well as Ari Cukierman, Brandon Hensley, the SPHEREx science team, and the participants in the workshop “Present and Future of Line Intensity Mapping” held at the Max Planck Institute for Astrophysics, Munich, for helpful discussions regarding this work. Y.-T.C. acknowledges the Balzan Cosmological Studies Travel Grant and the hospitality of Institut d'Astrophysique de Paris, where part of this work was conducted. K.W. acknowledges the support of the JPL SURF program. Y.-T.C. acknowledges support by NASA ROSES grant 18-2ADAP18-0192. B.D.W. acknowledges support by the ANR BIG4 project, grant ANR-16-CE23-0002 of the French Agence Nationale de la Recherche; the INFINITY NEXT project grant under the DIM ORIGINES Equipements 2023 program of the \^{I}le-de-France region; and the Simons Collaboration on ``Learning the Universe.'' The Flatiron Institute is supported by the Simons Foundation. T.-C.C. acknowledges support by NASA ROSES grant 21-ADAP21-0122. Part of this work was done at Jet Propulsion Laboratory, California Institute of Technology, under a contract with the National Aeronautics and Space Administration (80NM0018D0004). We acknowledge support from the SPHEREx project under a contract from the NASA/GODDARD Space Flight Center to the California Institute of Technology.

\software{
astropy \citep{2013A&A...558A..33A,2018AJ....156..123A}, COLOSSUS \citep{2018ApJS..239...35D},
ChainConsumer \citep{Hinton2016}
}

\appendix
\section{Line Intensity and Power Spectrum Derivations}\label{A:intensity_and_PS}

Here, we present a detailed derivation of the line intensity field and the window function of the power spectrum, as discussed in Sections~\ref{S:intensity} and \ref{S:Cl_nunu1_formalism}.

\subsection{Line Intensity Field}\label{A:line_intensity}
The intensity from the $i$th line at the angular position $\hat{n}$ and observed frequency $\nu$ is given by
\begin{equation}\label{E:nuInu_line_int}
\begin{split}
\nu I_\nu^i(\nu,\hat{n}) = \int d\chi\, \frac{d\left[ \nu_{\rm rf} L^i_{\nu}(\nu_{\rm rf},\chi, \hat{n})\right]}{dV}\frac{D_A^2(\chi)}{4\pi D_L^2(\chi)}.
\end{split}
\end{equation}
Here, $D_L$ is the luminosity distance, $D_A$ is the comoving angular diameter distance (equal to the comoving distance in a flat Universe), $\nu_{\rm rf}=\nu(1+z)$ is the rest-frame frequency corresponding to the observed frequency $\nu$ at redshift $z$, and $L^i_{\nu}(\nu_{\rm rf},\chi, \hat{n})=dL^i(\nu_{\rm rf}, \chi, \hat{n})/d\nu_{\rm rf}$ is the line profile (specific luminosity density) of the $i$th line at the angular position $\hat{n}$ and rest-frame frequency $\nu_{\rm rf}$. Note that $\nu_{\rm rf} L^i_{\nu}(\nu_{\rm rf},\chi, \hat{n}) = \nu L^i_{\nu}(\nu,\chi, \hat{n})$. 

The spectral resolution of typical LIM experiments cannot resolve the intrinsic line profile from sources. Therefore, we approximate $L^i_{\nu}(\nu_{\rm rf},\chi, \hat{n})$ as a Dirac delta function $\delta^D$ at the rest-frame line frequency $\nu^i_{\rm rf}$. We define the comoving line luminosity density:
\begin{equation}
M_{0,i}(\chi,\hat{n})=\frac{dL_i(\chi,\hat{n})}{dV} = \frac{d}{dV}\int d\nu_{\rm rf}\, L^i_{\nu}(\nu_{\rm rf},\chi, \hat{n}).
\end{equation}

With this, we get
\begin{equation}
\begin{split}
&\frac{d\left[ \nu_{\rm rf} L^i_{\nu}(\nu_{\rm rf},\chi, \hat{n})\right]}{dV} \\
&= M_{0,i}(\chi,\hat{n}) \nu_{\rm rf}\delta^D(\nu_{\rm rf} - \nu^i_{\rm rf})\\
&= M_{0,i}(\chi,\hat{n})\nu\delta^D(\nu - \frac{\nu^i_{\rm rf}}{1+z})\\
&= M_{0,i}(\chi,\hat{n})\nu\left.\frac{d\chi}{d\nu}\right|_{i\nu}\delta^D(\chi - \chi_{i\nu})\\
&= M_{0,i}(\chi,\hat{n})\frac{c(1+z_{i\nu})}{H(z_{i\nu})}\delta^D(\chi - \chi_{i\nu}).
\end{split}
\end{equation}
where $z_{i\nu} = \nu_{\rm rf}^i/\nu - 1$ is the redshift of the $i$th line at the observed frequency $\nu$, and $\chi_{i\nu}$ is its corresponding comoving distance. Here, $c$ is the speed of light, and $H(z)$ is the Hubble parameter.

Inserting this into Equation~\ref{E:nuInu_line_int}, we obtain
\begin{equation}\label{E:appendix_nuInu_line}
\nu I_\nu^i(\nu,\hat{n}) = \frac{c(1+z_{i\nu})}{H(z_{i\nu})} M_{0,i}(\chi_{i\nu},\hat{n})A_{0}(\chi_{i\nu}),
\end{equation}
where we define:
\begin{equation}
A_0(\chi) = \frac{D_A^2(\chi)}{4\pi D_L^2(\chi)}.
\end{equation}

\subsection{Window Function}\label{A:window_function}
The intensity field of the $i$th line in an observed filter with the filter profile function $R(\nu)$ is given by
\begin{equation}\label{E:nuInu_filt}
\langle\nu I_\nu^i(\nu)\rangle_{R} = \frac{\int d\nu' \,\nu' I_\nu(\nu')R(\nu')}{\int d\nu'
\,R(\nu')}.
\end{equation}
In this study, we consider the channel centered at a given observed frequency $\nu$ to be a top-hat function spanning $\nu^{\rm min} < \nu < \nu^{\rm max}$. Thus, the filter profile function is
\begin{equation}\label{E:tophat_filt}
R(\nu) = \frac{1}{\Delta \nu}\mathcal{I}(\nu; \nu^{\rm min}, \nu^{\rm max}),
\end{equation}
where $\Delta\nu=\nu^{\rm max}-\nu^{\rm min}$ is the filter width, and the indicator function $\mathcal{I}(\nu; \nu^{\rm min}, \nu^{\rm max})$ is defined by
\begin{equation}\label{E:indicator_function}
\mathcal{I}(\nu; \nu^{\rm min}, \nu^{\rm max})=
\begin{cases}
1 & \quad\text{if } \nu^{\rm min} < \nu < \nu^{\rm max}\\
0 & \quad\text{otherwise}.
\end{cases}
\end{equation}

The window function of the power spectrum at the channel centered at $\nu$, denoted as $W_{i\nu} (\chi)$ in Equation~\ref{E:Cl_line}, relates to the line intensity field by the following expression:
\begin{equation}
\langle\nu I_\nu^i(\nu)\rangle_{R} = \int d\chi\, W_{0,i\nu}(\chi),
\end{equation}
where we define $W_{0,i\nu}(\chi)=W_{i\nu}(\chi)/b_i(\chi)$. To obtain the window function for the power spectrum $W_{i\nu} (\chi)$, we first derive the expression of the intensity field in terms of the $\chi$ integration.

Inserting Equations~\ref{E:appendix_nuInu_line} and \ref{E:nuInu_filt} into Equation~\ref{E:tophat_filt}, we get
\begin{equation}
\begin{split}
\langle\nu I_\nu^i(\nu)\rangle_{R} &= \frac{1}{\Delta\nu}\int d\nu' \,\nu' I^i_\nu(\nu')\,\mathcal{I}(\nu'; \nu^{\rm min}, \nu^{\rm max})\\
&= \frac{1}{\Delta\nu}\int d\chi\, \nu'\frac{c(1+z)}{H(z)} M_{0,i}(\chi) A_{0}(\chi) \\
&\quad\quad\quad\quad\cdot\mathcal{I}(\chi; \chi_{i\nu}^{\rm min}, \chi_{i\nu}^{\rm max})\left.\frac{d\nu'}{d\chi}\right|_{i\nu'} \\
&= \int d\chi\,\frac{\nu'}{\Delta\nu} M_{0,i}(\chi) A_{0}(\chi)\,\mathcal{I}(\chi; \chi_{i\nu}^{\rm min}, \chi_{i\nu}^{\rm max}),
\end{split}
\end{equation}
where $\chi^{\rm min/max}_{i\nu} = \chi(z^{\rm min/max}_{i\nu})$, and $z^{\rm min/max}_{i\nu}=(\nu^i_{\rm rf}/\nu^{\rm min/max})-1$. Therefore, the window function of the $i$th line at the channel $\nu$ is
\begin{equation}
\begin{split}
W_{i\nu}(\chi) &= \frac{\nu}{\Delta\nu}b_{i}(\chi) M_{0,i}(\chi) A_{0}(\chi)\,\mathcal{I}(\chi; \chi_{i\nu}^{\rm min}, \chi_{i\nu}^{\rm max})\\
&=
\begin{cases}
\frac{\nu}{\Delta\nu}b_{i}(\chi) M_{0,i}(\chi) A_{0}(\chi)
& \quad\text{if } \chi^{\rm min}_{i\nu} < \chi < \chi^{\rm max}_{i\nu}\\
0 & \quad\text{otherwise}.
\end{cases}
\end{split}
\end{equation}

\subsection{Poisson Noise}\label{A:SN}
\begin{figure}[ht!]
\begin{center}
\includegraphics[width=\linewidth]{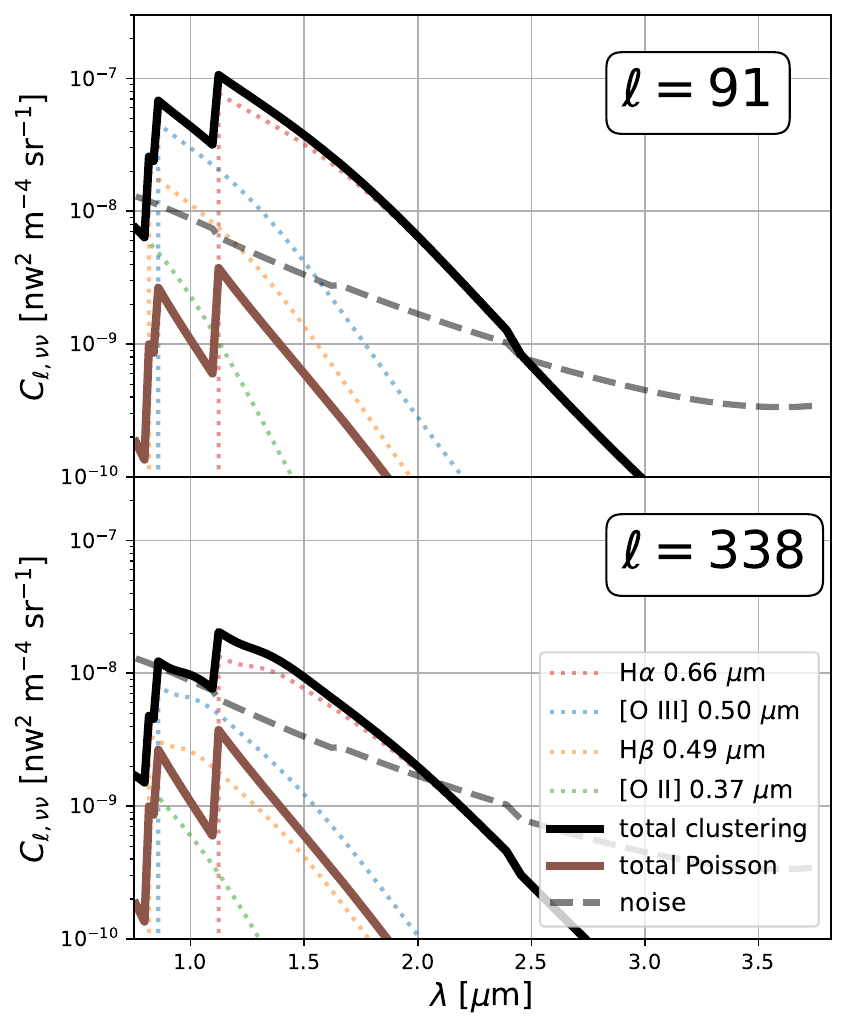}
\caption{\label{F:signal_model_Cl_SN} The clustering (black) and Poisson noise (brown) auto power spectrum from all four lines. The colored dotted lines denote the clustering terms of individual lines. The gray dashed line shows the SPHEREx noise power spectrum for reference. We show the power spectrum in the lowest- (top) and highest- (bottom) multipole modes in our fiducial setup.}
\end{center}
\end{figure}

The Poisson noise power from lines $i$ and $i'$ in channels $\nu$ and $\nu'$ is given by
\begin{equation}\label{E:ClP}
\begin{split}
C_{\ell,\nu\nu'ii'}^P =& \int d\chi\,D_A^2(\chi)\int dM \frac{dn}{dM}\\
&\cdot\frac{\nu L_\nu^i(M,\chi)}{4\pi D_L^2(\chi)}\frac{\nu' L_{\nu'}^{i'}(M,\chi)}{4\pi D_L^2(\chi)}\\
=& \frac{\nu}{\Delta\nu}\frac{\nu'}{\Delta\nu'}\int d\chi\,D_A^2(\chi)\left[\frac{1}{4\pi D_L^2(\chi)}\right]^2\\
&\cdot \int dM \frac{dn}{dM} L^i(M,\chi) L^{i'}(M,\chi),
\end{split}
\end{equation}
where $dn/dM$ is the halo mass function \citep{1999MNRAS.308..119S}. Here, we assume the line luminosities $L^i$ are functions of the halo mass $M$ and ignore scatters in this relation. The shot noise power depends on the details of the $L^i$--$M$ relation. Many previous studies have modeled this relation, ranging from scaling relations to semi-analytical models \citep[e.g.,][]{2016ApJ...817..169L,2017MNRAS.464.1948F,2022JCAP...02..026M}. Here, for simplicity, we assume a linear relation for $L^i$--$M$, which allows us to relate the last integral in Equation~\ref{E:ClP} to the luminosity density $M_{0,i}$ by
\begin{equation}
\begin{split}
\int & dM \frac{dn}{dM} L_\nu^i(M,\chi) L_{\nu'}^{i'}(M,\chi)\\
&=M_{0,i}(\chi)M_{0,i'}(\chi)\frac{\int dM \frac{dm}{dM}M^2}{\left(\int dM \frac{dm}{dM}M\right)^2}.
\end{split}
\end{equation}

Figure~\ref{F:signal_model_Cl_SN} compares the clustering and Poisson noise of the line power spectrum, as well as the SPHEREx noise level, in our fiducial case at our lowest- and highest-multipole bins. For our chosen scales and redshift range, the Poisson noise is much lower than the clustering signal and the noise. We also checked that the inclusion of Poisson noise has negligible effects on inference, and thus we ignore Poisson noise in our analysis.

\section{Redshift and Multipole Ranges}\label{A:ellmax}
Our choice of $z_{\mathrm{min}} = 0.7$ in Section~\ref{S:survey_setup} is based on the expected point-source sensitivity depth of SPHEREx. We estimate that below this redshift, we can reliably detect and constrain the redshift of the majority of galaxies with SPHEREx, and thus mask them to improve the sensitivity on diffuse line emission from fainter sources.

\begin{figure}[ht!]
\begin{center}
\includegraphics[width=\linewidth]{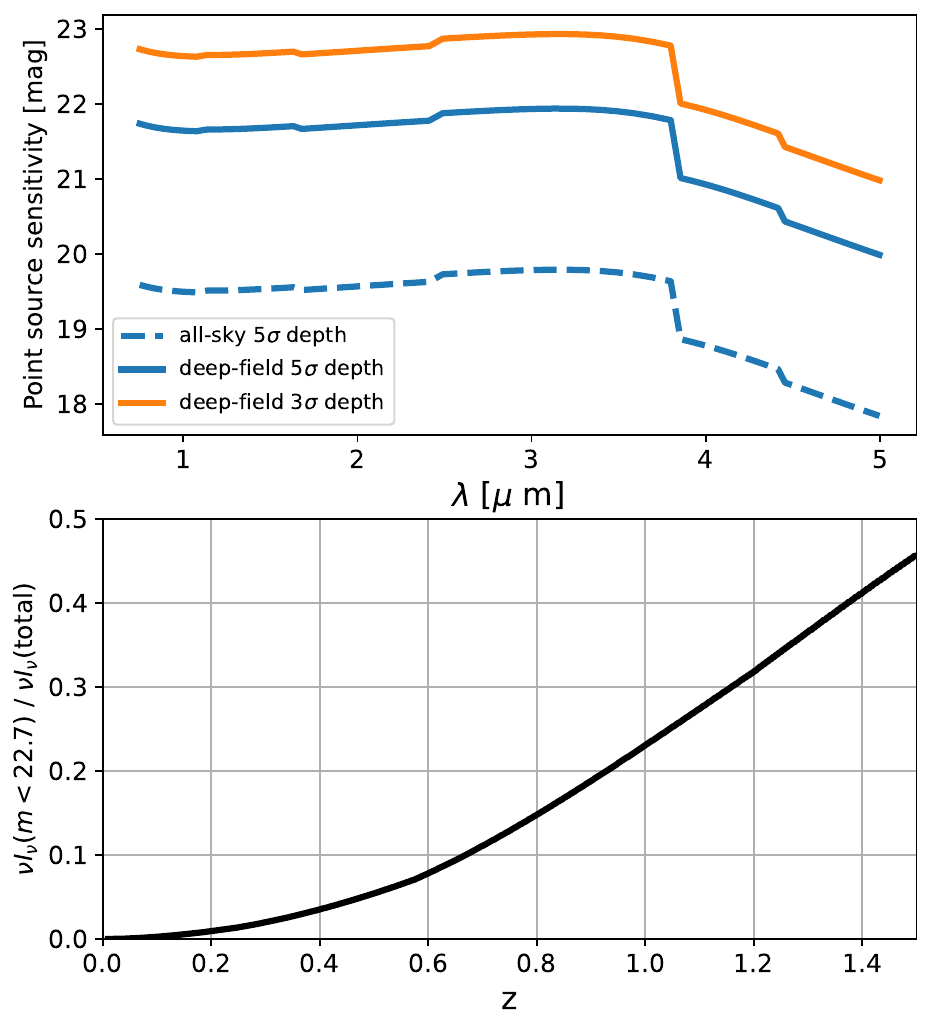}
\caption{\label{F:spherex_PSdepth}Top: the $5\sigma$ SPHEREx point-source sensitivity per channel on all-sky (blue dashed) and in deep fields (blue solid) in AB magnitudes. The orange line denotes the $3\sigma$ sensitivity per channel in deep fields, which is the depth we assumed for the point-source masking limit. Bottom: the fraction of total galaxy intensity below $m=22.7$ at $2\mu$m as a function of redshift.}
\end{center}
\end{figure}

SPHEREx is expected to achieve a $5\sigma$ point-source sensitivity of $m \sim 21.6$ per channel (at $\lambda < 3.82$ $\mu$m) in the deep fields\footnote{\url{https://github.com/SPHEREx/Public-products/blob/master/Point_Source_Sensitivity_v28_base_cbe.txt}} (the blue solid line in the top panel of Figure~\ref{F:spherex_PSdepth}). Considering that the SPHEREx photometric redshift fitting can achieve high accuracy for galaxies with $\gtrsim 3\sigma$ per channel sensitivity, this corresponds to a masking depth of $\sim 22.7$ at $2$ $\mu$m (the orange solid line in the top panel of Figure~\ref{F:spherex_PSdepth}). Using a model of the galaxy luminosity function across redshift and wavelength from \citet{2012ApJ...752..113H}, we estimate that below $z \sim 0.7$, there is $\lesssim 10\%$ of the integrated galaxy emission from sources below the masking depth (the bottom panel of Figure~\ref{F:spherex_PSdepth}), and thus we ignore any galaxy emission at $z < 0.7$ in our model.

\begin{figure}[ht!]
\begin{center}
\includegraphics[width=\linewidth]{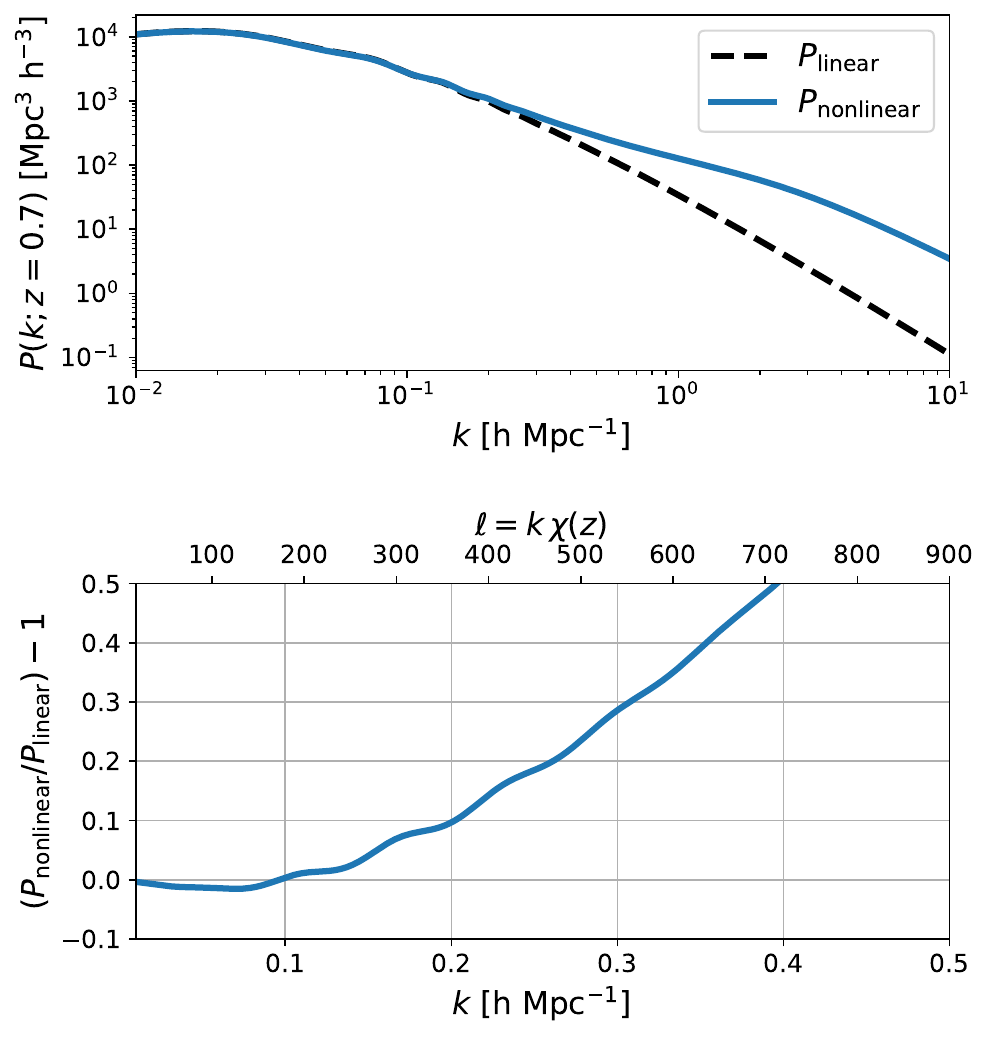}
\caption{\label{F:Pnl_z07}Top: linear (black dashed) and nonlinear (blue) matter power spectrum at $z=0.7$. The nonlinear power spectrum is obtained with the \texttt{syren-halofit} package \citep{2024A&A...686A.209B,2024A&A...686A.150B}. Bottom: the fractional difference between the linear and nonlinear power spectrum. The angular multipole mode $\ell$, corresponding to the transverse $k$ mode, is marked at the top axis. At $z=0.7$, the nonlinear clustering power deviates from the linear power by $\gtrsim10\%$ at $k\sim0.2$, corresponding to $\ell\sim350$. Therefore, we choose $\ell_{\rm max}=350$ for our analysis.}
\end{center}
\end{figure}

The choice of the maximum-$\ell$ mode ($\ell_{\rm max}=350$) is made to restrict our analysis to linear clustering scales. At $z_{\rm min}=0.7$, the effects of nonlinear clustering enter at $k\gtrsim 0.2$ $h$ Mpc$^{-1}$ (Figure~\ref{F:Pnl_z07}), and thus we set $\ell_{\rm max}$ to correspond to a transverse comoving maximum-$k$ mode of $0.2$ $h$ Mpc$^{-1}$ at $z_{\rm min}=0.7$ ($k_{\rm max}\sim \ell_{\rm max}/\chi(z_{\rm min})\sim 0.2$).

\section{Parameter Transformation}\label{A:parameter_transformation}
In this section, we present the relationship between the three sets of parameters: $\{c_{ij}\}$, $\{m_{ij}\}$, and $\{\theta_{ij}\}$. We will first describe the transformation between these parameters and then present the formalism for expressing the likelihood gradient and the Fisher matrix on one set of parameters in terms of another set of parameters.

The relationship between $\{c_{ij}\}$ and $\{m_{ij}\}$ is
\begin{equation}
m_{ij}=M_i(z_j+1) = \sum_{j=1}^{N_m}c_{ij}\hat{M}_j(z_j+1),
\end{equation}
where $\{z_j\}=\{-1,0,...,5\}$, and $\{\hat{M}_j(z)\}$ are the ReLU functions defined in Equation~\ref{E:M_basis_relu}. Representing $c_{ij}$ and $m_{ij}$ as the $N_{m}$-sized vectors $\mathbf{c}_i$ and $\mathbf{m}_i$, respectively, these two vectors follow a linear transformation relation defined by the Jacobian matrix $\mathbf{J}_i^{cm}$:
\begin{equation}
\mathbf{c}_i = \mathbf{J}_i^{cm}\mathbf{m}_i,
\end{equation}
and 
\begin{equation}
\mathbf{m}_i = \left(\mathbf{J}_i^{cm}\right)^{-1}\mathbf{c}_i,
\end{equation}
where 
\begin{equation}
\mathbf{J}_i^{cm}=
\begin{bmatrix}
1 & 0 & 0 & \dots && 0 \\
-2 & 1 & 0 & \dots && 0 \\
1 & -2 & 1 & \dots && 0 \\
0 & 1 & -2 & \ddots && \vdots \\
\vdots & \vdots & \vdots & \ddots & 1 &0 \\
0 & \dots & 0 & 1 & -2 & 1
\end{bmatrix},
\end{equation}
and 
\begin{equation}
\left(\mathbf{J}^{cm}_i\right)^{-1}=
\begin{bmatrix}
1 & 0 & 0 & \dots && 0 \\
2 & 1 & 0 & \dots && 0 \\
3 & 2 & 1 & \dots && 0 \\
\vdots & \vdots & \vdots & \ddots && \vdots \\
 & \dots & 3 & 2 & 1 & 0 \\
 & \dots & 4 & 3 & 2 & 1
\end{bmatrix}.
\end{equation}

The parameter set ${\theta_{ij}}$ is defined by the logarithm of ${m_{ij}}$:
\begin{equation}
\theta_{ij} = {\rm log} (m_{ij}),
\end{equation}
and thus the Jacobian $\mathbf{J}_i^{\theta m}$ is a diagonal matrix, with the diagonal elements given by $\mathbf{m}_i$:
\begin{equation}
\mathbf{J}_i^{m\theta} = {\rm Diag.}(\mathbf{m}_i).
\end{equation}
It is important to note that this transformation is nonlinear, and unlike $\mathbf{J}_i^{cm}$, the Jacobian $\mathbf{J}_i^{\theta m}$ depends on the specific parameter values.

During parameter inference, we concatenate parameters for all lines ($i$'s) into a single vector. In this case, the corresponding Jacobian matrix $\mathbf{J}$ is a block-diagonal matrix, with each block representing the Jacobian for each line ($\mathbf{J}_i$).

With the Jacobian matrix between the parameter sets, we can transform the likelihood derivatives and Fisher matrices between different parameter sets using the following relations:
\begin{equation}
\nabla^\beta {\rm log} \mathcal{L} = \mathbf{J}^{\alpha\beta} \nabla^\alpha {\rm log} \mathcal{L},
\end{equation}
and
\begin{equation}
\mathbf{F}^{\beta} = \left(\mathbf{J}^{\alpha\beta}\right)^T\mathbf{F}^{\alpha}\mathbf{J}^{\alpha\beta},
\end{equation}
where $\alpha$ and $\beta$ represent any two of the parameter sets ($\{c_{ij}\}$, $\{m_{ij}\}$, and $\{\theta_{ij}\}$). Note that the Jacobian matrices follow the chain rule: $\mathbf{J}^{\alpha\beta}=\mathbf{J}^{\alpha\gamma}\mathbf{J}^{\gamma\beta}$. Therefore, for example, we can obtain $\mathbf{J}^{c\theta}$ by $\mathbf{J}^{c\theta}=\mathbf{J}^{cm}\mathbf{J}^{m\theta}$.

In our algorithm, we initially compute the likelihood derivative and Fisher matrix in $\{c_{ij}\}$, then transform them to $\{\theta_{ij}\}$ during the Newton--Raphson optimization. Subsequently, we perform another transformation to $\{m_{ij}\}$ to quantify the model constraints.

\section{3D Power Spectrum Multipoles}\label{A:Pell}
Here, we provide a detailed expression for the interloper power spectrum multipoles, following the prescription from \citet{2021PhRvD.103f3523B}.

The $\ell$th multipole of the 3D power spectrum from interloper line $i$ projected to the target line $t$ is given by (Equation~\ref{E:Pell_i}):
\begin{equation}
\begin{split}
P_\ell^i(k,z_t) &= \frac{1}{\left(q^i_\perp\right)^2 q^i_\parallel} \,\frac{2\ell+1}{2}\\
&\cdot\int_{-1}^{1} d\mu \,P^i(k_i,\mu_i,z_i)\, \mathcal{L}_\ell(\mu),
\end{split}
\end{equation}
where $z_i=\lambda_t(1+z_t)/\lambda_i-1$ is the redshift of the $i$th line that contaminates the target line signal at the same observed frequency. $\mathcal{L}_\ell$ is the Legendre polynomial, and $q^i_\perp$ and $q^i_\parallel$ are the transverse and LOS projecting factors from the interloper redshift $z_i$ to the target line redshift $z_t$, respectively, given by
\begin{equation}
\begin{split}
    q^i_\perp &= \frac{D_A(z_i)}{D_A(z_t)},\\
    q^i_\parallel &= \frac{(1+z_i)/D_H(z_i)}{(1+z_t)D_H(z_t)},
\end{split}
\end{equation}
where $D_A$ is the comoving angular diameter distance, and $H(z)$ is the Hubble parameter. The intrinsic line power spectrum is given by
\begin{equation}
\begin{split}
    P^i(k_i,\mu_i,z_i) &= \left(1+\frac{f_i}{b_i}\mu_i^2\right)^2W(k_i,\mu_i, z)\\
    \cdot & b_i^2(z_i) \left(\nu I_{\nu,i}(z_i)\right)^2 D^2(z_i) P(k_i).
\end{split}
\end{equation}
Here, we only consider the clustering power on large scales, where the power spectrum is proportional to the square of the bias-weighted intensity $b_i(z)\nu I_{\nu,i}(z_i)$. $D(z)$ is the linear growth rate, and $P(k)$ is the linear power spectrum at the present day. The first term introduces intrinsic anisotropy from the RSD effect \citep{1987MNRAS.227....1K}, where $f_i\approx \Omega_m^{0.55}(z_i)$ is the linear growth rate at $z_i$. The corresponding Fourier mode and cosine angle of the interloper, $k_i$ and $\mu_i$, respectively, are given by
\begin{equation}
\begin{split}
k_i &= \frac{k}{q^i_\perp} \left[ 1 + \mu^2 \left( \left(F^i_{\rm proj}\right)^{-2} - 1 \right) \right]^{1/2},\\
\mu_i &= \frac{\mu}{F^i_{\rm proj}} \left[ 1 + \mu^2 \left( \left(F^i_{\rm proj}\right)^{-2} - 1\right) \right]^{-1/2},
\end{split}
\end{equation}
where $F^i_{\rm proj}=q^i_\parallel/q^i_\perp$.

The window function $W(k,\mu,z)$ is given by
\begin{equation}
W(k,\mu,z)={\rm exp}\left\{-k^2\left[\sigma_\perp^2(1-\mu^2)+\sigma_\parallel^2\mu^2 \right]\right\},
\end{equation}
where
\begin{equation}
\begin{split}
\sigma_\parallel &= \frac{c(1+z)}{H(z)R},\\
\sigma_\perp &= D_A(z)\sigma_{\rm beam},
\end{split}
\end{equation}
and $R$ is the spectral resolution, $D_A$ is the comoving angular diameter distance, and $\sigma_{\rm beam}$ is the beam size that we take for the SPHEREx pixel size of $6''.2$.

We note that a more comprehensive 3D power spectrum modeling requires accounting for the redshift evolution of signals along the LOS, which is beyond the scope of this study.

\bibliography{LIMlightcone}{}
\bibliographystyle{aasjournal}

\end{document}